\documentclass[namedreferences]{solarphysics}
\usepackage[hyperref,optionalrh,solaromanenum]{spr-sola-addons} % For Solar Physics 
\usepackage{graphicx}                    % For eps figures, newer &
\usepackage{float}
\usepackage{color}                       % For color text: \color
\newcommand{\aap}{    {\it Astron. Astrophys.}}

\newcommand{\apjl}{   {\it Astrophys. J. Lett.}}
\newcommand{\apjs}{   {\it Astrophys. J. Supp. Series}}

\begin{document}

\begin{article}

\begin{opening}

\title{New Insights into White-Light Flare Emission from Radiative-Hydrodynamic Modeling of a Chromospheric Condensation.}

%%%%%%%%%%%%%%%%%%%%%%%%%%%%%%%%%%%%%%%%%%%%%%%%%%%
%% Authors Names
%
\author{\surname{Adam~F.~Kowalski}$^{1,2}$\sep
       ~\surname{S.~L.~Hawley}$^{3}$\sep
        ~\surname{M.~Carlsson}$^{4}$\sep
        ~\surname{J.~C.~Allred}$^{2}$\sep   
        ~\surname{H.~Uitenbroek}$^{5}$\sep
~\surname{R.~A.~Osten}$^{6}$\sep
~\surname{G.~Holman}$^{2}$\sep
       }

%%%%%%%%%%%%%%%%%%%%%%%%%%%%%%%%%%%%%%%%%%%%%%%%%%%
%% Runningheads
%
\runningauthor{A.~F.~Kowalski \emph{et al.}}
\runningtitle{White-Light Flare Emission from a Chromospheric Condensation}

%%%%%%%%%%%%%%%%%%%%%%%%%%%%%%%%%%%%%%%%%%%%%%%%%%%
%% Affilations 
%
  \institute{$^{1}$Department of Astronomy, University of Maryland, College Park, MD 20742, USA.
                     email: \url{adam.f.kowalski@nasa.gov} \\ 
             $^{2}$NASA
  Goddard Space Flight Center, Heliophysics Science Division, Code 671, 8800 Greenbelt Rd., Greenbelt, MD  20771, USA. \\
  $^{3}$Department of Astronomy, University of Washington, Box 351580, Seattle, WA 98195, USA. \\
  $^{4}$Institute of Theoretical Astrophysics, University of Oslo,
  P.O. Box 1029, Blindern, N-0315, Oslo, Norway. \\
  $^{5}$National Solar Observatory, Sacramento Peak, P.O. Box 62, Sunpsot, NM 88349, USA. \\
  $^{6}$Space Telescope Science Institute, 3700 San Martin Drive, Baltimore, MD 21218, USA.
             }

%%%%%%%%%%%%%%%%%%%%%%%%%%%%%%%%%%%%%%%%%%%%%%%%%%%
%%% Abstract 
\begin{abstract}
The heating mechanism at high densities during M-dwarf flares is poorly understood.
Spectra of M-dwarf flares in the optical and near-ultraviolet
wavelength regimes have revealed three continuum components during the
impulsive phase:  1) an energetically dominant blackbody component with a
color temperature of $T\approx10^4$ K in the blue-optical, 2) a smaller amount of Balmer
continuum emission in the near-ultraviolet at $\lambda \le 3646$\ \AA\ and 3) an apparent pseudo-continuum
of blended high-order Balmer lines between $\lambda=3646$\ \AA\ and
$\lambda\approx3900$\ \AA.  These properties are not reproduced by models that employ a 
typical “solar-type” flare heating level of $\le 10^{11}$ erg cm$^{-2}$ s$^{-1}$ in non-thermal electrons, and therefore our understanding of these spectra is limited to a 
phenomenological three-component interpretation.  
  We present a new 1D radiative-hydrodynamic model of an M-dwarf flare
from precipitating non-thermal electrons with a large energy flux of $10^{13}$ erg cm$^{-2}$ s$^{-1}$.  The simulation produces bright near-ultraviolet and optical continuum
emission from a dense ($n>$10$^{15}$ cm$^{-3}$), hot ($T \approx
12\,000-13\,500$ K) chromospheric condensation.  For the first time, the
observed color temperature and Balmer jump ratio are produced self-consistently in a
radiative-hydrodynamic flare model.  We find 
that a $T\approx10^4$ K blackbody-like continuum component and a small Balmer jump
ratio result from optically thick Balmer ($\infty \rightarrow n=2$) and Paschen recombination ($\infty \rightarrow n=3$) radiation, and thus the properties of the  
flux spectrum are caused by blue ($\lambda\approx4300$\ \AA) light escaping
over a larger physical depth range compared to red ($\lambda
\approx6700$\ \AA) and near-ultraviolet ($\lambda\approx3500$\ \AA) light.  To model the near-ultraviolet pseudo-continuum
previously attributed to overlapping Balmer lines, we include the extra Balmer continuum opacity 
from Landau-Zener transitions that result from merged, high order energy levels
of hydrogen in a dense, partially ionized atmosphere.  This reveals a 
new diagnostic of ambient charge density in the densest regions of the atmosphere that are heated during dMe and solar flares.  

\end{abstract}

%%%%%%%%%%%%%%%%%%%%%%%%%%%%%%%%%%%%%%%%%%%%%%%%%%%
%% Keywords
%
%\keywords{Solar and Stellar Flares: Observations, Simulations, and Synergies.  Guest Editors:  Lyndsay Fletcher and Petr Heinzel.}

\end{opening}
%-------------------------------------------------

%%%%%%%%%%%%%%%%%%%%%%%%%%%%%%%%%%%%%%%%%%%%%%%%%%%
%% Sections
%
% \section{}%\label{s:?} 

\section{Introduction}

 It is notoriously difficult to explain the origin of heating at high atmospheric densities
 during solar and stellar flares.  
 The broadband color (white-light) distribution of the ultraviolet and optical emission
 during large and small flares on chromospherically active M-dwarf (dMe) stars 
 exhibits the general shape of a $T=8500-9500$ K blackbody \citep{Hawley1992,Hawley2003} without
 an indication of a significant Balmer jump at $\lambda=3646$\ \AA.  A recent homogeneous
 analysis of flare spectra around the Balmer jump has confirmed the presence of a hot
 blackbody (or blackbody-like) component with a color temperature of $T\gtrsim 10,000$ K that contributes
  most of the radiated near-ultraviolet (NUV) and optical energy in the impulsive phase \citep[][hereafter, K13]{HawleyPettersen1991, Kowalski2013}. 
 Intriguingly, the presence of 
 hot blackbody emission with $T\approx9000$ K has recently also been inferred from Sun-as-a-star observations
 during large (X-class) and small (C-class) solar flares \citep{Kretzschmar2011}.
 The hot (hereafter, ``hot'' refers to temperatures between 8500 and 15,000 K) 
 blackbody continuum component is often interpreted as evidence 
 of impulsive heating in a region with a high density of $n_e \approx10^{15}$ cm$^{-3}$ or more, which has been
 inferred from phenomenological modeling of dMe flares \citep{CramWoods1982, Houdebine1992, Christian2003}.  A recent 
 observation that directly supports this idea is the Vega-like
 flare spectrum detected during the 2009 January 16, megaflare event 
 on the dM4.5e star YZ CMi \citep{Kowalski2010, Kowalski2013}, which indicates that photospheric-level
 densities can be produced and maintained at sufficiently high temperatures during dMe flares.  However,
 heating high densities requires a large amount of volumetric energy deposition such that the 
 heating per unit mass is non-negligible.  Direct heating of the photosphere by non-thermal (NT) deka-keV electrons is 
 problematic because the electrons  
 experience rapid energy loss as soon as they impact the top of the ``thick-target'' chromosphere.

 A possible mechanism for increasing the density in the upper chromosphere was proposed in the early gas dynamic
simulations of \cite{Livshits1981}.  These authors showed that a shock from precipitating non-thermal electrons 
(with an energy flux of $10^{12}$ erg cm$^{-2}$ s$^{-1}$) produces a rapid downflow and compression of the material 
  to a very high ($>10^{15}$ cm$^{-3}$) density.  The development of downward moving ``chromospheric condensations'' (hereafter, CC) is accompanied by an upward moving ``evaporation'' of chromospheric material as it is heated beyond a million degrees \citep{Fisher1985}. Chromospheric evaporation and CCs are thought to be the fundamental processes in solar and stellar flares driving the 
frequently observed Neupert effect \citep{DennisZarro1993, Hawley1995, Gudel1996}.  Several studies have explored the role of explosive  CCs as the 
 source of white-light emission in solar and dMe flares \citep{Livshits1981, Gan1992, Katsova1997}, but 
 the formation threshold of CCs is strongly dependent on the initial atmospheric structure, the energy distribution (power law index, low-energy cutoff, and flux)
 of the precipitating NT electrons \citep{Fisher1989}, and the time-dependent ionization of helium \citep{Abbett1999, Allred2005, Allred2006},
 thus necessitating an accurate and realistic description of the NT electron distribution, the pre-flare atmosphere structure, and the 
 level populations of important cooling agents.  

  Recent radiative-hydrodynamic (RHD) simulations of dMe \citep{Allred2006} flares with the RADYN code \citep{Carlsson1995, Carlsson1997} explored the atmospheric
response to 
 an actual NT electron spectrum that was obtained from the peak of an X-class solar flare inferred from \emph{Reuven Ramaty High Energy Solar Spectroscopic Imager} (RHESSI) data \citep{Holman2003}.
  A constant, solar-type flare heating level of the NT electrons was employed with injected energy fluxes of $10^{10}$ erg cm$^{-2}$ s$^{-1}$ (F10) and 
 $10^{11}$ erg cm$^{-2}$ s$^{-1}$ (F11). 
 These simulations produced explosive mass motions indicative of 
 chromospheric evaporation and a CC, but with densities far too low to produce
 strong, blackbody-like white-light emission.  Moreover, the prominent spectral component in these simulations was 
 a hydrogen recombination spectrum with strong 
 Balmer recombination continuum emission and a smaller amount of Paschen recombination continuum emission.
 In these models, the heating at high density was achieved through Balmer and Paschen continuum backwarming of the 
 photosphere, which increased in temperature by 1200 K in the F11 run and produced additional amounts of optical/red (blackbody-like) emission.
  For the range of non-thermal electron energy fluxes thus far explored in RHD models \citep[F9\,--\,F11; see also][]{HawleyFisher1994, Abbett1999}, a flare spectrum with hot blackbody emission has not been produced through the development of a CC, through photospheric heating from Balmer and Paschen continuum backwarming \citep{Allred2006}, or through photospheric heating from X-ray and EUV backwarming \citep[][see also the extensive comparison of RHD model predictions and recent observations in Section 9 of \cite{Kowalski2013}]{Hawley1992, Allred2006}.

 In addition to hot blackbody emission, the recent spectral observations of dMe 
flares also show evidence for a secondary continuum component in the NUV at $\lambda \lesssim 3700$\ \AA.  
 A jump in flux at these wavelengths
 is not accounted for by an extrapolation of a single hot blackbody component that can be fit to the redder wavelengths at $\lambda >4000$\ \AA. 
 The jump in flux relative to the blackbody flux is smaller in the peak phase than in the gradual decay phase, when the color temperature 
 of the hot blackbody component also decreases.  A comparison to the F11 RHD model spectra from \cite{Allred2006}
 led to the conclusion that Balmer recombination continuum 
 emission was needed to account for the additional flux in the NUV \citep{Kowalski2010}. 
  However, high resolution ($R\approx40\,000$) echelle flare spectra 
 show that the NUV continuum from $\lambda \le 3700$\ \AA\ can exhibit the color temperature of
 a hot blackbody ($\approx11\,000$ K) while not exhibiting any signature of a Balmer discontinuity or edge feature  \citep{Fuhrmeister2008} that is indicative 
 of Balmer recombination continuum emission.  Instead, the highest order Balmer 
 lines are broadened and appear to diminish as they converge into a ``pseudo-continuum'' at wavelengths between $\lambda  = 3646$\ \AA\ 
 and the bluest identifiable Balmer line, which is typically H15 $\lambda$3712 or H16 $\lambda$3704 for both large solar and dMe 
flares \citep{Donati1984, HawleyPettersen1991, Kowalski2010}.

The broadening and merging of the high order Balmer lines in flares is thought to be due to the well-known Stark effect from ambient
charges \citep{Svestka1963, Donati1985}.
 The Balmer edge itself is also thought to be broadened, but thermodynamically self-consistent (where the internal partition function converges)
modeling of this effect was not available until the occupational probability theory of pressure ionization was developed by \cite{HM88}.  
This improved description of pressure ionization from Stark broadening at the Balmer edge is currently employed in white dwarf
atmospheric modeling codes \citep{Tremblay2009}, but has not yet been implemented in models of solar and stellar flares
to determine if the theory can explain the NUV/blue continuum, and in particular, the lack of a Balmer edge.  
The Stark broadening of the higher order Balmer lines is 
potentially a useful tool for constraining the environment (\emph{i.e.}, charge density) where
 the (blackbody) continuum emission is produced in stellar flares \citep{CramWoods1982}.

What atmospheric conditions and atomic processes at high density produce a flare radiation field that self-consistently exhibits all the important continuum components and line properties in the blue and NUV?  First, we must relate the observables in blue/NUV spectra to physical parameters of the atmosphere. Then, the spectra can be used as powerful tools to constrain the heating mechanisms that produce these atmospheric conditions.  In this paper, we explore the possible role of a very dense CC for producing white-light emission during dMe flares, which reveals a meaningful interpretation of the spectra that will be used to constrain targeted modeling of individual solar and stellar flares. 
  These simulations are timely due to new spectral constraints in the blue/NUV and optical wavelength regimes \citep{Kowalski2013} 
and modern computational facilities enabling calculation of detailed spectra with sophisticated
radiative transfer codes.  
Section \ref{sec:method} describes our method of solution and the starting atmospheric conditions, Section \ref{sec:blackbody} describes 
the atmospheric dynamics and a new interpretation of the $10^4$ K blackbody color temperature, Section \ref{sec:nonideal} describes the improved modeling of the Balmer edge, Section \ref{sec:real} explores the viability of the model, Section \ref{sec:discussion} summarizes the study and  presents several conclusions, and Section \ref{sec:work} describes future modeling work.

\section{Method of Solution and Initial Atmosphere} \label{sec:method}
\subsection{NLTE Radiative-Hydrodynamic Modeling with RADYN}
We performed 1D radiative-hydrodynamic modeling using the RADYN code 
\citep{CarlssonStein1992, Carlsson1994, Carlsson1995, Carlsson1997, Carlsson2002}.  RADYN solves
the equations of mass, momentum, internal energy, non-equilibrium level populations, charge conservation, and radiative transfer \citep[see also][]{Abbett1999}. 
 The electron and ion
temperatures are equal and an adaptive grid \citep{Dorfi1987} with 250 grid points resolves shocks in the atmosphere.  The NLTE (non-LTE) problem is solved in RADYN using the technique of \cite{Scharmer1981} and \cite{Scharmer1985}. 
The NLTE level populations for three atoms are calculated:
 a hydrogen atom (six levels including H \textsc{ii}), a helium atom (five levels for He \textsc{i}, three levels for He \textsc{ii}, and He \textsc{iii}), and a singly ionized
calcium ion (six levels including the Ca \textsc{iii} ground state), giving a total of 22 bound-bound (b-b) and 19 bound-free (b-f) transitions calculated in detail \citep[see][]{Allred2005}. 
Continua from elements
other than H, He, and Ca \textsc{ii} are treated as background continua in LTE using the Uppsala atmospheres program
\citep{Gustafsson1973}.
A radiative cooling function from CHIANTI \citep{Dere1997, Young2003} at 
42 temperature points between $10^4$ K and $10^8$ K is included to account for optically thin cooling.  
 Radiation from  several ionization states 
 (\emph{e.g.}, Mg \textsc{ii}, Fe \textsc{ii}, and Si \textsc{ii}) that are likely optically thick at $10^4$ K (Hansteen, 2012, private communication) may be important in flares \citep{Avrett1986}
but is not yet included in the models (see Section \ref{sec:work}).

Each line profile is calculated with up to 100 frequency points across the transition
 using a Voigt profile and complete redistribution.  For Lyman transitions, the effects
of partial redistribution are approximated by truncating the line
profiles at ten Doppler widths.
A change to the line broadening due to the Stark effect is implemented in RADYN compared to the 
models of \cite{Allred2006}.  Here, we use
the analytic formulae (Method 1) from \cite{Sutton1978} to add the Stark damping parameter to the total
damping parameter in the Voigt profiles for hydrogen.  The Stark broadening of calcium and helium transitions remains the same as in previous versions of the code. 

Backwarming from coronal X-ray and extreme-ultraviolet (XEUV) radiation 
is treated according to the method of \cite{HawleyFisher1994}, \cite{GanFang1990}, and \cite{Abbett1999},
which approximates the radiation source using a plane-parallel geometry within 1 Mm 
and as a point-source beyond this distance.  
The emissivity at 41 temperature points from $10^4 - 10^8$ K in 14 wavelength bins from $\lambda=1-2500$\ \AA,
is generated from the Astrophysical Plasma Emission Code \citep[APEC;][]{SmithAPEC}.  The hydrogen, helium, and metal cross sections
are obtained from CHIANTI. 

We ran flare simulations with the starting atmospheric parameters and boundary conditions similar to those used
 in \cite{Allred2006}, in order to facilitate the comparison.   
A preflare, plane-parallel 1D atmosphere with log $g =$ 4.75 was relaxed to have velocities $\lesssim10$ m s$^{-1}$
and converged NLTE populations.  The upper temperature boundary was initially fixed at $6 \times 10^{6}$ K ($n_e \approx 2.6\times 10^{10}$ cm$^{-3}$), and
constant non-radiative heating was applied at column mass $m$ greater than 3.16 g cm$^{-2}$ ($\tau_{5000}=0.06$) 
to emulate a convective energy flux.  The atmosphere was relaxed using a
 plane-parallel approximation of the X-ray backwarming, as would be appropriate
from extended overlying active region loops. The effective temperature of the relaxed atmosphere 
is 3630 K, corresponding to that of an early to mid M-dwarf.
 The height coordinate ($z$) of the atmosphere lies 
along the length of a semi-circular magnetic
loop of half-length $L_{1/2} = 10^{9}$ cm (10 Mm), 
with $z=0$ at the photosphere ($\tau_{5000}=1.0$, log $m =$ 1.14 g cm$^{-2}$).
A modification to the gravitational acceleration
is incorporated near the top of the loop \citep{Abbett1999}; otherwise, the geometry is assumed to be a
vertical tube of constant cross-sectional area. 
 The imposed 1D geometry assumes the plasma motion is confined to the
direction of the magnetic field, as would be appropriate for a low plasma $\beta$ (plasma $\beta=\frac{P_{\rm{gas}}}{P_{\rm{mag}}}$), but is not necessarily 
 satisfied in the photosphere of the preflare atmosphere (the equipartition magnetic field strength is 4.2 kG at the photosphere, consistent with observations \citep{JKV1996}) or in the chromosphere during the flare. 
 Detailed molecular
transitions that are important for flux redistribution in an M-dwarf photosphere have not yet been included
in the opacity.

\subsection{Flare Heating Prescription}
We model the atmospheric heating from a double power law, non-thermal (hereafter, NT)
 electron beam injected at the top of the atmosphere.  In this study, we seek to directly connect to the work of \cite{Allred2006},
which used the thick-target parameters inferred from RHESSI \citep{Lin2002} hard X-ray spectra
during the peak of the 2002 July 23 X4.8 GOES class solar flare \citep[][see also \cite{Ireland2013}]{Holman2003}.
A double power law of electron energies is parameterized by
$\delta_l = 3.0$ (the power law index from the cutoff energy to the break energy),
$\delta_u = 4.0$ (the power law index at energies greater than the break
energy), $E_c = 37$ keV (the ``cutoff energy'' or lowest energy of the NT electron spectrum), 
and $E_B = 105$ keV (the break energy between $\delta_l$ and $\delta_u$). The value of $E_c$ derived from observations 
is an upper limit, and $E_c=37$ keV is moderately large compared to typically assumed or inferred values on the order of 20 keV.
Note also that the break in hard X-ray photon spectra (and thus also the inferred double power law shape of the NT electron spectrum) can be the result of 
the non-uniform ionization of the ambient atmosphere \citep{Su2011} or from energy losses to the return-current electric field 
\citep{Holman2012}. 
 The parameters of the NT electron spectrum
are poorly constrained during dMe flares because the hard X-ray emission is too faint to be observed except during the largest flares \citep{Osten2010}.  
A powerful superflare on the RS CVn HR 1099 is consistent with a hard NT electron spectrum with a single power law index $\delta\approx3$ \citep{Osten2007},
similar to the relatively hard double power law spectrum employed in our simulations.

Free parameters which may not be well-constrained by RHESSI observations alone are the 
energy flux \citep{Krucker2011} and the precise time profile (duration) of the flare heating in 
any given magnetic loop.  We extend the range of heating fluxes from \cite{Allred2006} to also
consider those with $10^{12}$ erg cm$^{-2}$ s$^{-1}$ (F12) and $10^{13}$ erg cm$^{-2}$ s$^{-1}$ (F13), in addition
to $10^{11}$ erg cm$^{-2}$ s$^{-1}$ (F11).  The F13 
  flux is not much larger than the largest values that have been derived during solar flares \citep{Neidig1994}, and such a level has recently
 been suggested to be relevant to the brightest solar flare kernels \citep{Krucker2011, Milligan2014, Gritsyk2014}.  The duration of the heating in a single
 magnetic loop in solar flares is constrained in some cases to $\le 25-60$~s \citep{Wang2007, Schrijver2006} and in other cases may be several minutes \citep{Qiu2010, Warren2006}.  A wavelet analysis of high-time resolution hard X-ray light curves for 647 solar flares observed with the \emph{Compton Gamma Ray Observatory} (CGRO)
 indicates a range in the duration of individual bursts from 0.1~s to 50~s \citep{Aschwanden1998}.
 The shortest burst durations from
 high-time resolution constraints observed in the NUV in dMe flares range from $\approx2$~s \citep{Robinson1995} to 10~s \citep{Mathioudakis2006}.  For our heating profile, we use a constant energy flux for a duration of 2.3~s, after which the NT electron energy deposition decreases to 0,
 and the cooling 
phase is modeled up to 5.1~s for the F13 model (and 30~s for the models with lower beam fluxes).  Short heating bursts
represent the heating of a single loop, which can be superposed in sequence to emulate an arcade development as has 
been done for soft X-ray data of solar flares \citep{Warren2006}.

The elastic collisions between NT electrons and the ambient charged and neutral components of the
atmosphere heat the gas through Coulomb interactions.
Following \cite{HawleyFisher1994}, \cite{Abbett1999}, and
\cite{Allred2005,Allred2006}, the NT electron energy deposition rate is included as an extra
 term in the conservation of energy equation. 
The Coulomb heating rate as a function of height 
 is modeled here using the analytic, non-relativistic test-particle approach of 
\cite{Emslie1978}, which was modified to include the height-dependent ionization fraction of the atmosphere in \cite{HawleyFisher1994}.
This prescription is based on the collisional thick-target model \citep{Brown1971},
but it also includes an approximation for pitch angle scattering of the NT electrons 
\citep{Emslie1978}. 
 The inelastic collisions between NT electrons and the ambient neutral 
hydrogen component of the atmosphere cause direct excitation and ionization \citep{MottMassey}; 
we follow \cite{Kasparova2009} and add these NT collisional rates to the thermal rates 
for hydrogen according to the prescription in \cite{Fang1993}\footnote{A comprehensive study of NT 
rates was presented in the 
calculations of \cite{Ricchiazzi1982} and \cite{Ricchiazzi1983}, 
who found that the H \textsc{i} $n=1 \rightarrow \infty$ and $n=1 \rightarrow n=2$ rates are important, 
but that the $n=2 \rightarrow \infty$ rate was negligible compared to the $n=2 \rightarrow \infty$ thermal rates for
their range of beam fluxes.}.  The results from \cite{DalganoGriffing} are included in these rates 
to account for secondary ionizations of hydrogen by electrons that are
ionized by the beam from the ground state of hydrogen.

\section{Properties of the NUV and Optical Continuum Predictions}  \label{sec:blackbody}
\subsection{Comparison to Observables} \label{sec:blackbody_31}
Continuum predictions from radiative-transfer modeling are constrained 
with two simple observational parameters:  1) the Balmer jump ratio and 2) 
the color temperature
of the continuum at wavelengths longer than the Balmer edge.  These two parameters
were referred to as $\chi_{\mathrm{flare}}$ and $T_{\mathrm{BB}}$, respectively, in a recent 
homogeneous analysis of 20 flares on dMe stars from K13. A blackbody function is fit to the blue window (BW) wavelength ranges (see K13) of the flux spectrum from $\lambda=4000-4800$\ \AA\ (the blue-optical wavelength regime) to obtain $T_{\mathrm{BB}}$, while $\chi_{\mathrm{flare}}$ is the ratio of the flux averaged over 30 \AA\ centered on $\lambda=3615$\ \AA\ (blueward of the Balmer jump) to the flux averaged over 30 \AA\ centered on $\lambda=4170$\ \AA\ (redward of the Balmer jump).  These wavelength ranges are least affected by emission lines while also being near the Balmer jump (see K13).  We note that the detected continuum emission shortward of the Balmer jump in K13 also included 
 wavelengths between $\lambda=3420$ \AA\ and $\lambda = 3600$ \AA.  A more secure detection and characterization of this continuum 
emission would include shorter wavelengths \citep[such as in][for a solar flare observed from space]{Heinzel2014}, but the signal-to-noise decreases at $\lambda < 3600$ \AA\ due to the atmospheric cutoff.  
The distribution of these parameters was found to fall between $\chi_{\mathrm{flare}}=1.5-4$ 
and $T_{\mathrm{BB}}=9000 - 14\,000$ K during the peak phases of nearly all of the flares in the sample from K13.  
The value of $\chi_{\mathrm{flare,peak}}$ varied with flare type, with the most impulsive flares 
exhibiting a narrow range of $1.5-2.2$ and the gradual flares exhibiting larger values.  
We measure these same parameters from the flux spectra for the peak phases of the 
model continua for the range of beam fluxes studied here.  The model continua are calculated 
at a fixed number of points across the NUV and optical, and so we interpolate the continua to the wavelengths
used in the observations.  
 We also compute the ratio of the line-integrated, continuum-subtracted
H$\gamma$ emission line flux to the value of the continuum flux at $\lambda=4170$\ \AA\ (hereafter, H$\gamma_{\rm{EW4170}}$; note, this was referred to as H$\gamma$/C4170 in K13).
  
The model predictions for $\chi_{\rm{flare}}$, $T_{\rm{BB}}$, H$\gamma_{\rm{EW4170}}$ are given for each NT electron flux in Table \ref{table:observables}.  For the F13 simulation, the table gives the coarse time evolution
of these parameters for the impulsive and gradual phases.  The evolution of the flux spectrum for F13 is shown in 
Figure \ref{fig:cont_evol}\footnote{The flux spectra for the F13 simulation are available upon request.}.  
At early times, the F13 flux spectrum exhibits a large Balmer jump ratio and a cool color temperature in the blue-optical.
However, from $t=1.2$~s to $2.2$~s, the continuum at $\lambda > 3646$\ \AA\ becomes bluer 
as the flux at $\lambda=4300$\ \AA\ increases significantly while the 
flux at $\lambda \approx 6690$\ \AA\ increases only marginally.  This results in an increase in the blue-optical color temperature, $T_{\mathrm{BB}}$, of 2000 K.  Also, the Balmer jump ratio decreases from 3.3 to 2 as the flux 
 at $\lambda \approx 3550$\ \AA\ decreases while the flux at $\lambda=4300$ increases.  The continuum 
at wavelengths shorter than the Balmer edge also becomes 
bluer over this time interval as the flux at $\lambda<3000$\ \AA\ increases while the flux at $\lambda \approx 3550$\ \AA\ decreases.  In Section \ref{sec:origin}, we will explain how 
these changes in the flux spectrum are related to the increasing hydrogen b-f opacities in the atmosphere.

At $t=2.2$~s, the F13 simulation produces a flux spectrum that falls into the regime 
of the observations for the peak phase of dMe flares with 
a relatively small Balmer jump ratio, $\chi_{\mathrm{flare}}=$ 2.0, and a blue-optical color temperature of $T_{\mathrm{BB}}=9300$ K. The F13 model NUV flux between $\lambda = 3000-3646$\ \AA\ also exhibits the color temperature of a hot blackbody with $T=11\,500$ K. 
Although the signal-to-noise of spectra at NUV wavelengths is lower than the data at blue-optical wavelengths (due to the atmospheric cutoff), the color temperature in the NUV is generally consistent with the value derived at blue-optical wavelengths (\emph{cf.} Figures $8-9$ of K13).   
 The H$\gamma_{\rm{EW4170}}$ also falls within the range of the observed values at flare peak (\emph{cf.} Figures $11-12$ of K13).  Blue-optical
emission with a color temperature higher than $\approx$6000 K has not previously been self-consistently reproduced in RHD models of dMe flares (K13).
The F11 and F12 simulations do not produce properties that are consistent with observations of the impulsive phase:
the Balmer jump ratios are too large ($>7$) and the color temperatures are too low ($<6000$ K).  From Table \ref{table:observables}, it can be seen that the time-evolution of the characteristics of the F13 simulation
is consistent with the general time-evolution observed in dMe flares:  the Balmer jump and H$\gamma_{\rm{EW4170}}$
 decrease in the rise phase and increase in the decay phase.  The timescale of the 
burst (5~s), however, is much shorter than the timescales of most dMe flares (minutes to hours).  Adding up sequentially heated
magnetic strands as in the spatial development of a flare arcade \citep[\emph{e.g.},][]{Warren2006, Qiu2010} 
is outside the scope of this work, but we intend to pursue this in a future paper.

\subsection{Dynamics and Energy Balance in the F13 Atmosphere} \label{sec:dynamics}
The evolution of the F13 flux spectrum in Figure \ref{fig:cont_evol} 
is tied to the dynamical evolution of the atmosphere.  
We discuss here general properties of the hydrodynamics.

The NT electrons initially deposit their energy over a broad extent, with significant 
volumetric heating from 275 km (log $m =-2.1$ g cm$^{-2}$) to the top of the chromosphere at 450 km (log $m =-4.05$ g cm$^{-2}$), and 
there is non-negligible NT electron energy deposition ($300$ erg s$^{-1}$ cm$^{-3}$) directly in the corona.  
The maximum energy deposition from the NT electrons ($8\times10^5$ erg s$^{-1}$ cm$^{-3}$)
initially occurs at $z_o=390$ km (log $m =-3.05$ g cm$^{-2}$), 
which is 99\% neutral but with a significant 
electron density $n_e=10^{12}$ cm$^{-3}$.   Most of the NT energy is deposited initially 
into NT hydrogen ionization at $z_o$.  As the electron density increases at $z_o$
 an increasing
fraction of the NT energy goes into raising the temperature since heating of electrons
occurs more efficiently than heating of neutrals (note, the thermalization of electrons, protons, and
neutrals is assumed to happen instantaneously in these models).  

Hydrogen becomes 
 completely ionized by $t=0.002$~s and thermal 
ionization increases over NT ionization as the temperature exceeds $20\,000$ K.  He \textsc{i} 
then becomes the primary cooling agent as the temperature reaches $40\,000$ K.  At $t=0.003$~s, the temperature approaches  
65\,000 K and most helium
at this height is now singly ionized.  From $t=0.003$~s to $t=0.01$~s
 the He \textsc{iii} fraction increases from 2\% to 99.6\% at $z_X=425$ km as the temperature rapidly increases  
 to 140\,000 K.  At $t=0.016$ sec, a localized temperature maximum of 180\,000 K 
forms at $z_X$, and the thermal pressure
generates upward velocities of 4 km s$^{-1}$ and downward velocities of 2 km s$^{-1}$.  
These mass motions are the seeds to the explosive condensation and evaporation which follow from
runaway heating as material is evacuated from this region of the atmosphere. 
  
Meanwhile at $t=0.016$~s, between $z=450$ and $z=460$ km (at the preflare transition region heights), 
there is a strong upward flow of 30 km s$^{-1}$ resulting from the increase 
in temperature to 90\,000 K of lower density material (previously the upper chromospheric material)
 between the site of maximum beam heating (425 km) and the transition
region (450 km).  At $t=0.2$~s, the evaporation consists of two fronts: a high density chromospheric front with maximum velocity of 125 km s$^{-1}$ and 
a low density transition region front with maximum velocity of 250 km s$^{-1}$.  The downward front is a high density ``chromospheric condensation'' (CC), 
with a maximum velocity of 95 km s$^{-1}$.  The maximum densities of the chromospheric fronts have 
increased from $n_H \approx 2\times10^{14}$ cm$^{-3}$ ($t=0.2$~s) to $n_H \approx 5\times10^{14}$ cm$^{-3}$ ($t=0.4$~s) and the temperatures have decreased from $50\,000$ K
to $37\,000$ K.  At the location of $z_X$, the temperature has reached nearly 10 MK by $t=0.4$~s from the runaway NT electron heating as the density has decreased (material is evacuated) from $n_H = 5\times10^{13}$ cm$^{-3}$ (t$=0$~s) to $n_H = 6\times10^{12}$ cm$^{-3}$ ($t=0.4$~s). At the upper and lower ends of the transition from very hot ($T \approx 400\,000$ K$ - 10$ MK) plasma to $T\approx35\,000$ K plasma, 
upward and downward propagating \emph{shock fronts} are located at the very steep (nearly discontinuous) jumps in gas pressure, where the He \textsc{iii} fraction changes from 100\% to $<30$\%.\footnote{The speed of sound in the CC at $t=0.4$~s is 30 km s$^{-1}$ whereas material is traveling at speeds of nearly 100 km s$^{-1}$.}
 
At $t=0.33$~s, the timesteps of the simulation become unmanageably small ($\Delta t \approx10-100$ ns) due to the 
precision requirements of RADYN in calculating the population levels of He \textsc{i} and He \textsc{ii} in regions
of strong temperature gradient, especially at the upper evaporation fronts.  
At $t=0.478$~s, we decrease the accuracy of the minor level populations, with
accurate solutions provided only for level population densities that are $> 10^{-9}$ of the total elemental population density
(initially, the parameter is $10^{-12}$), 
while the second derivative of the grid weights is increased to weight more toward the higher density (CC) front.  
The result is a dramatic increase in the time step size until $t=5.1$~s as the low-density (upper) chromospheric evaporation front
is smoothed over (the number of grid points here drastically decreases and the temperature is no longer maintained
at $35\,000 - 50\,000$ K).  The results for the simulation
at $z > 450$ km are thus less accurate, but this approximation is required in order to compute 1D simulations with very steep and dynamic 
temperature and density gradients. 
We focus only on the radiating material in the 
CC and below, since this higher density material produces most of the optical emission; the model above 450 km should not affect the analysis at the lower heights.  

 Figure \ref{fig:dynamics} summarizes the dynamical
evolution of the lower atmosphere at $t=2.2$~s, when the maximum white-light emission 
is present. The top panel shows the velocity of the downward moving CC.
At this time in the CC, the maximum hydrogen density is $7.6\times10^{15}$ cm$^{-3}$, or 17.4$\times10^{-9}$ g cm$^{-3}$, which is over a factor of 12
greater than the hydrogen density at this height ($z=255.4$ km) before the flare.  This mass density
is comparable to the preflare density at $z\approx145$ km, which lies between the temperature minimum and the lower chromosphere.  The evolution of the 
speed and temperature of the maximum density zone in the CC is shown in the bottom panels.  
By $t=2.2$~s, the speed of the downflowing material has decreased to $\approx$50 km s$^{-1}$ as it encounters 
denser material and higher gas pressure lower in the atmosphere, the equation of motion for which was derived analytically by \cite{Fisher1989}.   
Over the simulation, the temperature of the region of maximum density
in the CC decreases steadily;  this is a result of larger hydrogen emissivity for lower temperatures while
 increasing the density in the CC gives a larger volumetric cooling rate.  
Eventually the temperature nearly settles between 12\,500 and 13\,000 K at $t=2.2$ s as continuous injection of NT electrons
 deposit energy in the CC.
After the NT electron energy deposition
decreases to zero at $t=2.3$~s, the CC continues to propagate downward, 
increasing its density to $1.3\times10^{16}$ cm$^{-3}$\ at $t=4.0$~s.  The temperature drops to 7000 K at this time (when the NT energy input is cut off), which is 
much larger than the preflare temperature of $\approx5000$ K at this height (196.5 km).

At $t=2.2$~s, the $E\approx 65-200$ keV NT electrons encounter a thick-target ``wall'' at the CC and produce a maximum
energy deposition rate of $5\times10^{7}$ erg cm$^{-3}$ s$^{-1}$ (compared to a maximum of $< 10^6$  erg cm$^{-3}$ s$^{-1}$ initially) at $z=256.9$ km; 
the full width at half maximum (FWHM) of the energy deposition rate is only 0.46 km (compared to 
75 km initially).  The simulation becomes computationally problematic shortly after this time and therefore
the NT heating rate is cut off at 2.3 s. 
 The energy budget around the CC is shown in Figure \ref{fig:energy}, which 
 gives the individual contributions to the change in total internal energy \emph{per} unit mass.  Positive quantities contribute toward
heating, excitation, and ionization; negative quantities contribute toward cooling, recombination, and de-excitation.  
The lower end of the CC from Figure \ref{fig:dynamics} at 239.4 km is indicated 
by a vertical long-dashed line in Figure \ref{fig:energy}.
In the CC, the NT electron energy deposition is balanced by the thin and thick (detailed) radiative losses, 
resulting in a net decrease in the internal energy (recombination and cooling).  At deeper layers 
than the CC, $E \gtrsim 200$ keV NT electrons can still penetrate and heat the atmosphere.  At $z\approx200$ km,
there is net ionization and heating from both Balmer continuum backwarming (red diamonds) and NT electron energy deposition.  At these depths, radiation from the optically thin loss function and 
from the detailed transitions does not counterbalance these sources of energy.  There are also some contributions from compressional and viscous work
where the CC impacts and ``accretes'' the stationary material below it.  

\subsection{Properties of the Contribution Function for Continuum Emission in the Flare Atmosphere} \label{sec:origin}
\begin{figure}[H]
\begin{center}
\includegraphics[scale=0.6]{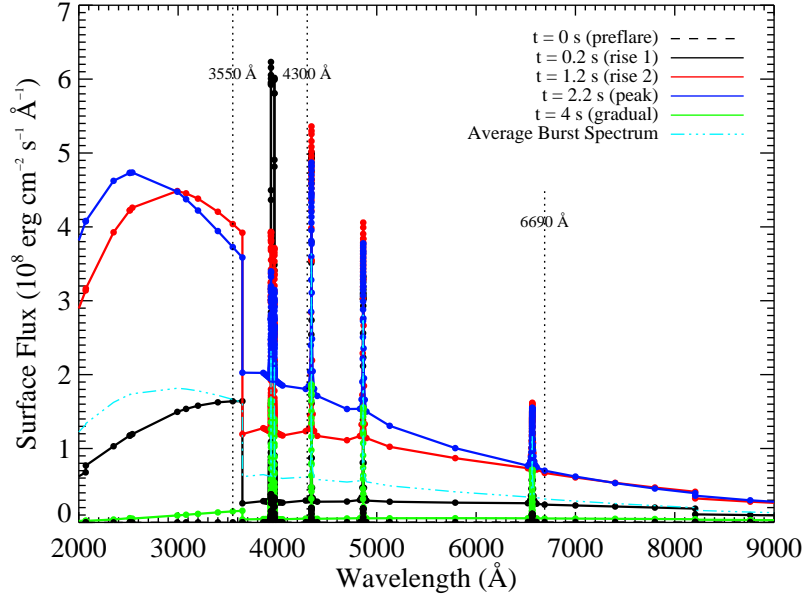}
\caption{The evolution of the NUV and optical flux spectra
during the F13 simulation.  The observable parameters extracted from these spectra are given in Table \ref{table:observables}. The coadded (average) burst spectrum is discussed in Section \ref{sec:real}.  The preflare spectrum is not visible on this scale.  The vertical dotted lines indicate the NUV ($\lambda=3550$ \AA), blue-optical ($\lambda=4300$ \AA), and red-optical ($\lambda=6690$ \AA) wavelengths used in the continuum analysis (Section \ref{sec:origin}).}
\label{fig:cont_evol}
\end{center}
\end{figure}

\begin{figure}[H]
\begin{center}
\includegraphics[scale=0.55]{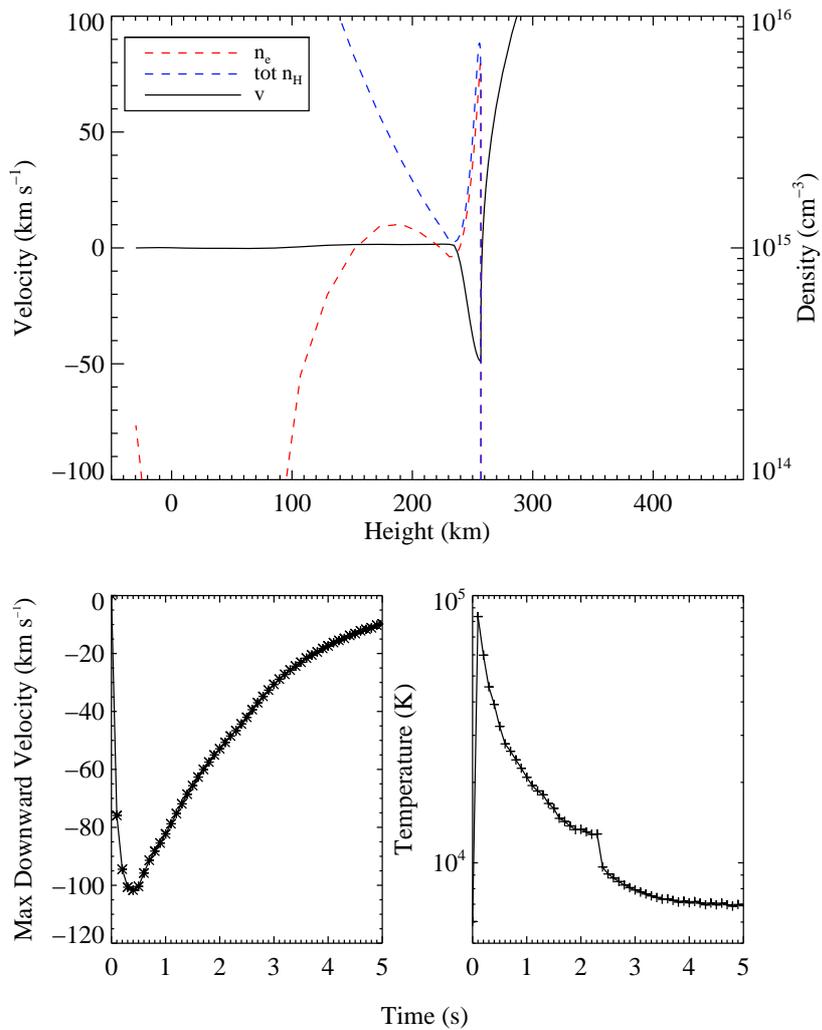}
\caption{\emph{Top:} The velocity field in the lower atmosphere at $t=2.2$~s shows a chromospheric condensation (CC)
with a large hydrogen and electron density exceeding $5\times10^{15}$ cm$^{-3}$. \emph{Bottom left:}
The time-dependence of the velocity in the highest density zone in the CC. \emph{Bottom right:}
The time-dependence of the temperature in the highest density zone in the CC. The velocity of the CC changes 
smoothly through the burst while the temperature reflects the cutoff of the NT electron energy deposition at $t=2.3$~s.
}
\label{fig:dynamics}
\end{center}
\end{figure}

\begin{figure}[H]
\begin{center}
\includegraphics[scale=0.4]{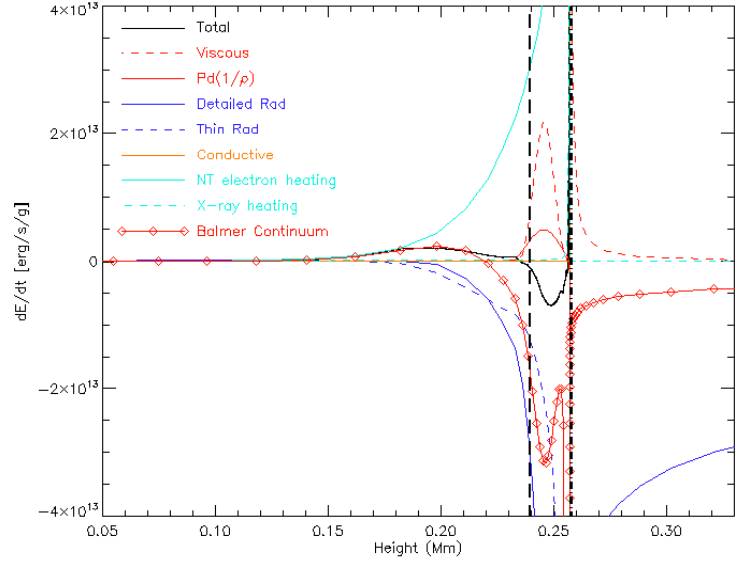}
\caption{Contributions to the energy balance for $t=2.2$~s in the F13 simulation.  These curves show the 
 \emph{net} amount for each component.  For the detailed radiation, the heating and cooling are integrated over the wavelengths
of the transitions.  The Balmer continuum (red diamonds) heating/cooling is shown as a separate curve 
from the total detailed radiation curve (which includes the Balmer continuum).  The CC, with velocities lower than 5 km s$^{-1}$,
is material between $z=239.4$ km (long dashed vertical black line) and 257.4 km (thick vertical dash-dotted line).
}
\label{fig:energy}
\end{center}
\end{figure}

The CC decreases in temperature and increases in density over time (Figure \ref{fig:dynamics}).
By $t=2.2$~s the CC has attained temperatures $>$12\,000 K and increased in density over 12 times the preflare density, thus 
producing a large amount of radiative cooling from hydrogen and other elements (Figure \ref{fig:energy}).  In this section, we 
explore the height dependent properties of the emergent intensity from the CC at three continuum wavelengths calculated in detail: 
 $\lambda=3550$\ \AA\ in the NUV, $\lambda=4300$\ \AA\ in the blue-optical, and $\lambda=6690$\ \AA\ in the red-optical, 
which are indicated in Figure \ref{fig:cont_evol}.  
In Section \ref{sec:blackbody_31}, we found that the NUV and blue-optical F13 model peak spectrum (blue spectrum in 
Figure \ref{fig:cont_evol}) is described by two blackbody spectra with color temperatures of $T=11,500$ K between $\lambda=3000-3646$\ \AA\ and $T=9300$ K at $\lambda>4000$\ \AA\ in relative amounts given by the value of the Balmer jump ratio ($\approx2$).
In this section, we use the contribution function to the emergent intensity \citep{Magain, Carlsson1998} to explain 
the origin and characteristics of the white-light continuum emission in this model spectrum.  
We use the following form of the contribution function, $C_I = dI_{\lambda}/dz$ \citep{Leenaarts2013}, for a plane-parallel
atmosphere\footnote{The differential $dI_{\lambda}/dz$ in the definition of the contribution function is 
the same as the differential describing the variation of the specific intensity with height in the radiative transfer equation; however, these differentials are not equivalent.}:  

\begin{equation} \label{eq:didz}
C_I = \frac{dI_{\lambda}}{dz} = \frac{\chi_{\nu}}{\mu} S_{\nu} e^{-\tau_{\nu}/\mu} \frac{c}{\lambda^2}
\end{equation}

\noindent where $\chi_{\nu}$ is the height-dependent 
extinction coefficient (in units of cm$^{-1}$; not to be confused with $\chi_{\rm{flare}}$ the Balmer jump ratio), 
$\tau_{\nu}$ is the optical depth ($\int_{z}^{10^9\rm{cm}} \chi_{\nu} dz$),
and $S_{\nu}$ is the height-dependent source function.  
More simply, this equation is $\frac{j_{\nu}}{\mu} e^{-\tau_{\nu}/\mu} \frac{c}{\lambda^2}$, or the plasma emissivity ($j_{\nu}$) at a given height
and frequency multiplied by the attenuation of the radiation (determined by $e^{-\tau_{\nu}}$) as it propagates outward in the direction of $\mu$
from that height.\footnote{$j_{\nu}$ is often used to describe
 the emissivity in units of photons cm$^{-3}$ s$^{-1}$ sr$^{-1}$ Hz$^{-1}$ whereas $\eta_{\nu}$ is used for the emissivity in units of ergs cm$^{-3}$ s$^{-1}$ sr$^{-1}$ Hz$^{-1}$; here, we use 
$j_{\nu}$ to describe the emissivity in units of ergs cm$^{-3}$ s$^{-1}$ sr$^{-1}$ Hz$^{-1}$.}
 We discuss the atomic processes that contribute to the continuum emissivity in Section \ref{sec:opaccontrib}.  

In Figure \ref{fig:didz1}, we show the contribution function to the emergent vertical intensity ($\mu \approx 1$) for 
$\lambda=3550$\ \AA, $4300$\ \AA, and $6690$\ \AA\ at $t=2.2$~s.  The temperature profile (dashed line) of the flare
atmosphere is qualitatively similar to that of the quiescent atmosphere, except that the flare chromosphere is hotter and has moved to
higher column mass.  
Radiation escapes from two principal regions of the flare atmosphere:  a stationary component (see Figure \ref{fig:dynamics}) produces a wide maximum in $C_I$ at $z\approx200 $ km in the mid-/upper- flare chromosphere below the CC, while the CC produces a very thin component peaking at $z\approx256.3$ km with maxima in $C_I$ of 950, 140, and 120 erg cm$^{-2}$ s$^{-1}$ \AA$^{-1}$ sr$^{-1}$ cm$^{-1}$ (extending above the top of the figure) for $\lambda=3550$, 4300, and 6690\ \AA, respectively.  Smaller values of $z$ (defined in Section \ref{sec:method}) refer to
larger physical depths, whereas larger values of $\Delta z$ refer to 
larger physical depth \emph{ranges} in the atmosphere.
We measure the FWHM of the narrow component of $C_I$ to compare the physical depth ranges ($\Delta z_{\rm{FWHM}}$) over which radiation escapes from the CC.
 The $\Delta z_{\rm{FWHM}}$ is the largest for $\lambda=4300$\ \AA: the $\Delta z_{\rm{FWHM}}$ values are 1 km, 1.75 km, and 1.25 km for $\lambda=3550$, 4300, and 6690\ \AA, respectively. 
The contribution function in the stationary component of the flare atmosphere is also markedly different
among these three wavelengths.  Whereas $\lambda=4300$\ \AA\ has a local maximum at 200 km, and $20-25$\% of the emergent intensity originates from layers below the CC at $z < 230$ km, the other two wavelengths have very little emergent intensity 
originating from these depths, with $\lambda=3550$\ \AA\ showing the least.  
The arrows on the top of the figure indicate the layers at which $\tau_{\lambda}=1$ for each wavelength.  The $\tau_{\lambda}=1$ layer occurs in the CC for $\lambda=3550$ and 6690\ \AA, and in the stationary flare layers for $\lambda=4300$\ \AA. 
Therefore, the intensity at $\lambda=4300$\ \AA\ originates over the largest physical depth range (largest $\Delta z$) and from the 
lowest atmospheric heights due to the smallest optical
depth in the CC.  

At times earlier than 2.2~s, a larger fraction of the emergent intensity at $\lambda=3550$\ \AA\ and 6690\ \AA\ originates from the stationary component of the flare atmosphere 
because the CC has a lower density and optical depth.
The evolution of the optical depth for $\lambda=4300$\ \AA\ and $3550$\ \AA\ and the evolution of the column densities ($N$) for hydrogen in $n=2$ and $n=3$
in the CC (where $v < -5 $ km s$^{-1}$) are shown in Figure \ref{fig:tau_evol}.  As the CC becomes denser and attains 
temperatures that excite hydrogen to $n=2$ and $n=3$, the probability of photoionization by light with these wavelengths increases.  
The values of the column densities at $t=2.2$~s of $N_{H}(n=2) > 3 \times10^{17}$ cm$^{-2}$ and $N_{H}(n=3) > 10^{17}$ cm$^{-2}$ define useful requirements\footnote{See Section \ref{sec:opaccontrib}, there are additional opacities
from photoionization of the higher levels of hydrogen, hydrogen free-free absorption, and opacity from the H$^-$ ion
that can also contribute towards the 
optical depth at these wavelengths.} for large optical depths ($\tau_{4300}=0.8, \tau_{3550}=5$) in any future flare model
of CCs with material near temperatures of 12\,000 K.

The evolution of the contribution function results from increasing, wavelength-dependent optical depths
 at optical/NUV wavelengths
and can explain the evolution of the flux spectrum in Figure \ref{fig:cont_evol} (Section \ref{sec:blackbody_31}).
The very large and continuously increasing optical depth at $\lambda=3550$\ \AA\ causes a \emph{decrease}
in the flux at this wavelength between $t=1.2$~s and $t=2.2$~s because the radiation can only escape from a very thin layer.  At $t=2.2$~s, the radiation at $\lambda=4300$\ \AA\ is still relatively optically thin in the CC ($\tau < 0.8$), 
and therefore has a higher probability of escape, giving rise to a larger FWHM of the contribution function in the CC ($\Delta z_{\rm{FWHM}}$) and also a larger fraction that 
can escape from the stationary component of the flare atmosphere.  
The flux at 4300\ \AA\ increases from $t=1.2$~s to $2.2$~s in Figure \ref{fig:cont_evol} and results
 from the increasing density but a lower optical 
depth due 
to the lower photoionization cross-section and smaller population density for $n=3$ compared to $n=2$.  The larger optical depth
at $\lambda=6690$\ \AA\ relative to $\lambda=4300$\ \AA\ is caused by a larger photoionization cross-section (but same population density),
which leads to a much smaller increase in flux at this wavelength
in Figure \ref{fig:cont_evol} from $t=1.2-2.2$~s.  The NUV flux significantly increases at $\lambda < 3000$\ \AA\ and the color temperature at $\lambda=3000-3646$\ \AA\ becomes larger 
over this time interval due to similar reasoning for the increase of the flux at $\lambda=4300$\ \AA\ and the increase of the color temperature at $\lambda > 4000$\ \AA.  Higher signal-to-noise flare spectra at $\lambda < 3600$\ \AA\ would help to rigorously test the slope changes predicted by these models at NUV wavelengths.

\subsection{Opacity Components in the Optical Depth} \label{sec:opaccontrib}
\subsubsection{The Optically Thick Hydrogen Recombination Flare Spectrum for the F13 model} \label{sec:thick}
\begin{figure}[H]
\begin{center}
\includegraphics[scale=0.55]{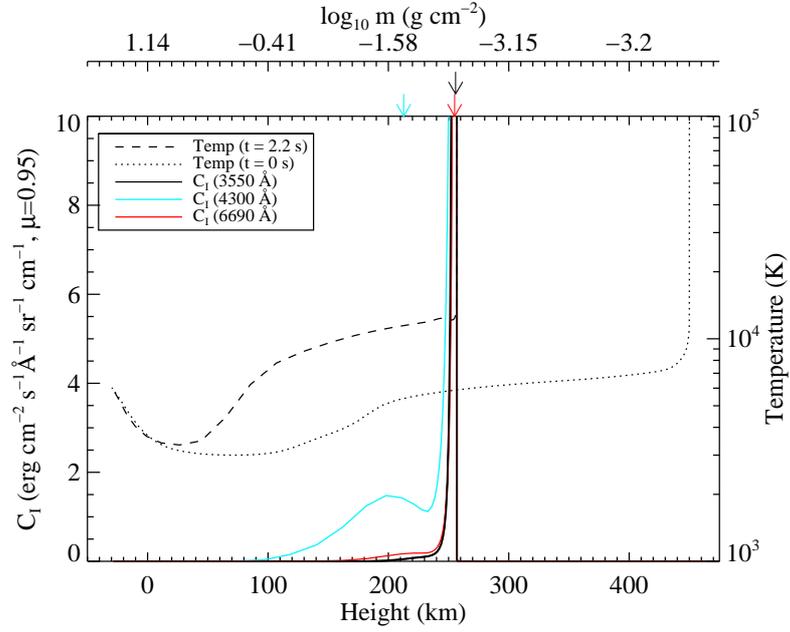}
\caption{ The contribution to the emergent intensity for three continuum wavelengths in the NUV, blue, and red at $t=2.2$~s compared to the 
temperature structure (for the same height range as in Figure \ref{fig:dynamics}).  The column mass scale for $t=2.2$~s is shown 
at the top axis.  The arrows indicate where $\tau=1$ occurs for each wavelength.  The contribution functions in the CC ($z \approx 239-257$ km)
are extremely thin and extend
off the top of the plot (see text).
}
\label{fig:didz1}
\end{center}
\end{figure}

\begin{figure}[H]
\begin{center}
\includegraphics[scale=0.6]{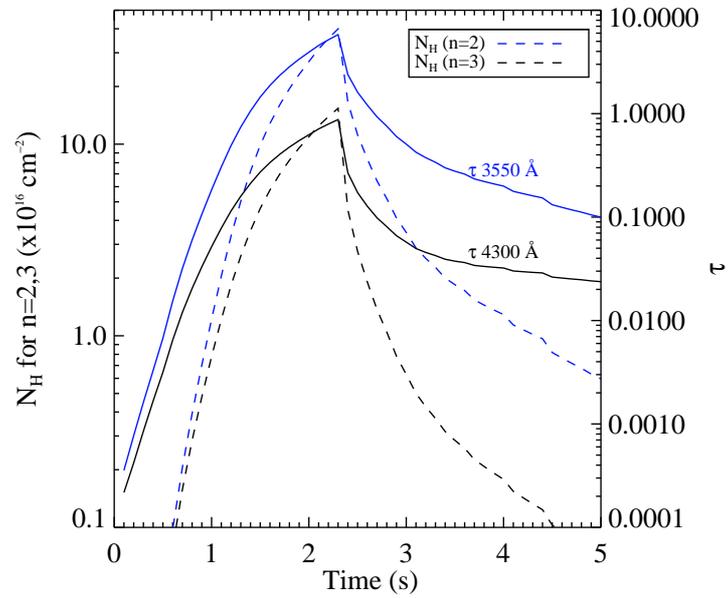}
\caption{The evolution of column density for $n=2$ and $n=3$ of hydrogen (left axis, dashed lines) and optical depth (right axis, solid lines) for atmospheric heights in the CC
with velocities $<-5$ km s$^{-1}$.  
At $t=2.2$~s, the CC has an extent of 18 km from 239 km to 257 km and has temperatures $>12\,000$ K.  
Approximately 3.5$\times10^{12}$ erg cm$^{-2}$ s$^{-1}$ of NT electron energy flux is being deposited within the CC at 
this time.
}
\label{fig:tau_evol}
\end{center}
\end{figure}

\begin{figure}[H]
\begin{center}
\includegraphics[scale=0.55]{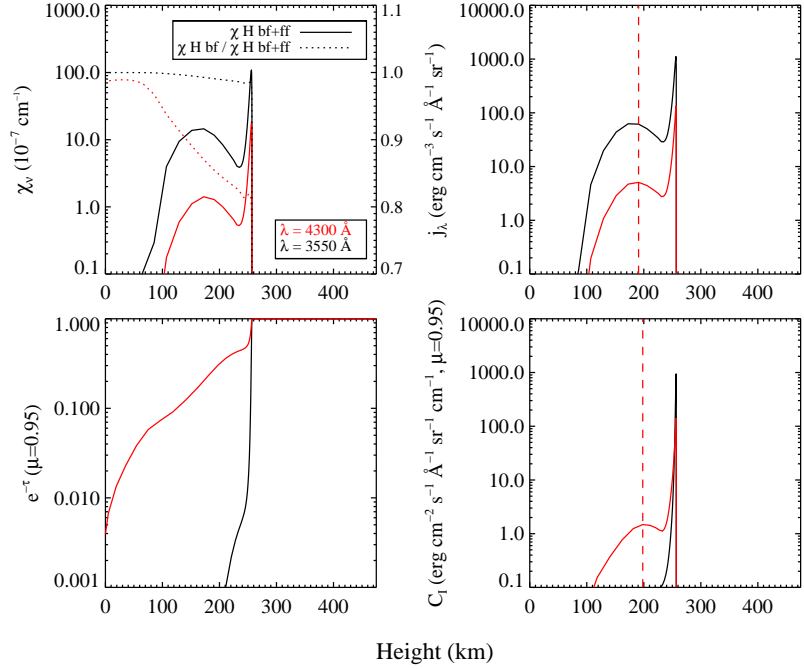}
\caption{ The contribution function (bottom right) for 4300\ \AA\ and 3550\ \AA\ is broken up into three 
parts: opacity (top left), emissivity (top right), and the attenuation (bottom left).  The ratios of hydrogen b-f opacity to total hydrogen (b-f and f-f) opacity are shown in the top left panel (right axis).  In the stationary flare layers ($z\approx190$ km), there is a 
large emissivity at $\lambda=3550$\ \AA\ but also a very large attenuation due to photoionizations of hydrogen from $n=2$ in the higher layers of the CC.  The peaks of the emissivity and contribution function in the stationary layers for $\lambda=4300$\ \AA\ are indicated by vertical dashed lines (see text).
}
\label{fig:didz2}
\end{center}
\end{figure}

In this section, we quantitatively demonstrate how the properties of the flux spectrum at $t=2.2$~s in Figure \ref{fig:cont_evol} primarily arise
 from spontaneous hydrogen recombination continuum emission by reconstructing the contribution function to the emergent intensity
from atomic emissivity processes.  The b-f opacity ($\chi_{\nu}$) corrected for stimulated emission is given by \citet[][Equation 7-1]{Mihalas1978}:

\begin{equation} \label{eq:bf_ideal}
\chi_{\nu} = \Sigma_i \{ [n_i - n_i^* e^{-h\nu/kT}]  \sigma_i(\nu) \}
\end{equation}

\noindent where $\sigma_i$ is the photoionization cross-section of an atom with an electron in level $i$ for light of frequency $\nu$;
 $n_i^* = n_c \times (n_i/n_c)^*$ where $n_c$ is the number density of protons and $(n_i/n_c)^*$ is the density ratio in LTE which can be calculated using the Saha equation.  
We calculate the hydrogen b-f and hydrogen free-free (f-f, bremsstrahlung) opacity at $\lambda=3550$\ \AA, 4300\ \AA, and 6690\ \AA\
for the flare atmosphere at $t=2.2$~s.  
The total hydrogen continuum (b-f$+$f-f) opacity values for $\lambda=4300$ and 3550\ \AA\ are shown in the top left of Figure \ref{fig:didz2}.
The hydrogen b-f opacity dominates for both wavelengths, 
but the hydrogen f-f is non-negligible in the CC at optical wavelengths.  At $\lambda=4300$\ \AA, the Paschen b-f ($n=3\rightarrow \infty$)
continuum opacity is largest, whereas at $\lambda=3550$\ \AA\ the Balmer b-f ($n=2 \rightarrow \infty$) continuum opacity dominates.  
The b-f and f-f opacities from the H$^-$ ion are not shown explicitly in Figure \ref{fig:didz2} but are included in RADYN and in the contribution
function (Figure \ref{fig:didz1}).  The H$^-$ opacity is small ($\approx$3.5\%) compared to the hydrogen b-f and f-f opacities in the CC but dominates at the lower temperatures ($T<9000$ K) in the flare atmosphere at $z \lesssim 130$ km.  Although the electron density is 
high in the CC, Thomson scattering is relatively very small.  

Most of the CC is so dense that the LTE condition $S=B$ (where $B$ is the Planck function) applies and the source function is only dependent on local quantities.  The level populations at $t=2.2$~s for $n=2$ and $n=3$ are equal to the LTE populations, and collisional rates are generally large compared to radiative rates, for most of the flare atmosphere ($z = 100 - 256.6$ km) except at the surface of the CC.  The level population $n_p$ is the LTE value through all of the CC.  The hydrogen b-f and f-f 
continuum emissivity, $j_{\lambda}$, is shown in the top right panel of Figure \ref{fig:didz2}.  For simplicity, only the spontaneous recombination emission is shown.  The relative rates of induced to spontaneous recombination are very small at these wavelengths but can become larger at redder wavelengths. The hydrogen emissivity can be approximated as $\{\chi_{\nu,b-f}+\chi_{\nu,f-f}\}B_{\nu}\frac{c}{\lambda^2}$ through most of the CC; but to account 
for departures from LTE at the surface of the CC, we use the NLTE hydrogen b-f emissivity from Equation 7-2 of \cite{Mihalas1978}.
The ratio of emissivities ($j_{\lambda=3550}/j_{\lambda=4300}$) in the CC is 
about 9.5, as we expect for hydrogen recombination (Balmer and Paschen) continuum emission at these temperatures.  
When $\tau$ is non-negligible, the wavelength dependent attenuation becomes
important in the contribution function (Equation (\ref{eq:didz})).
The attenuation factor, $e^{-\tau_{\lambda}}$, is shown in the bottom left panel of Figure \ref{fig:didz2}, and illustrates how 
the $\lambda=3550$\ \AA\ emissivity is preferentially attenuated at $z=180-200$ km, where $\tau_{3550}=20$.  Here, $\tau_{4300}\approx 1$ which is large enough to attenuate some of the radiation from these heights (compare the shift in peaks at this height indicated by vertical dashed lines in the top right and bottom right panels).  The $e^{-\tau_{\lambda}}$ 
factor gives rise to the wavelength-dependent changes in the FWHM of $C_I$ ($\Delta z_{\rm{FWHM}}$) and in the fraction of the emergent intensity that originates from the stationary component (Section \ref{sec:origin} and Figure \ref{fig:didz1}).

The contribution function, $C_I$, is shown again in the bottom right panel of Figure \ref{fig:didz2} (note logarithmic y-scale as compared
to Figure \ref{fig:didz1} which is on a linear scale).  Integrating $C_I$
over height gives the emergent intensity at $\mu=0.95$.
 The integration over height reveals that the Balmer jump intensity 
 ratio ($I_{\lambda=3550}/I_{\lambda=4300}$, similar to $\chi_{\mathrm{flare}}$) is $\approx$2.2, a value indicative
 of the small Balmer jump in emission in the flux spectrum in Figure \ref{fig:cont_evol}. 
 The emergent optical intensity ratio $I_{\lambda=4300}/I_{\lambda=6690}$ of 2.3 is a value consistent with the color temperature of 
 a hot ($\approx9000$ K) blackbody.  We have thus shown that hydrogen recombination (and bremsstrahlung) 
 continuum emissivity between $11\,000 - 14\,000$ K 
 integrated over heights where optical depths become large ($\tau_{4300} \gtrsim 0.5$) 
and are wavelength dependent (from the hydrogen cross-section dependence and thermal population density variation)
 can self-consistently generate the color temperature ($T_{\rm{BB}}$) of 
the blue-optical emission and the Balmer jump ratio ($\chi_{\rm{flare}}$) in agreement with observations of 
the peak phases of dMe flares.  
 
\subsubsection{The Optically Thin Hydrogen Recombination Flare Spectrum for the F11 model}

\begin{figure}[H]
\begin{center}
\includegraphics[scale=0.4]{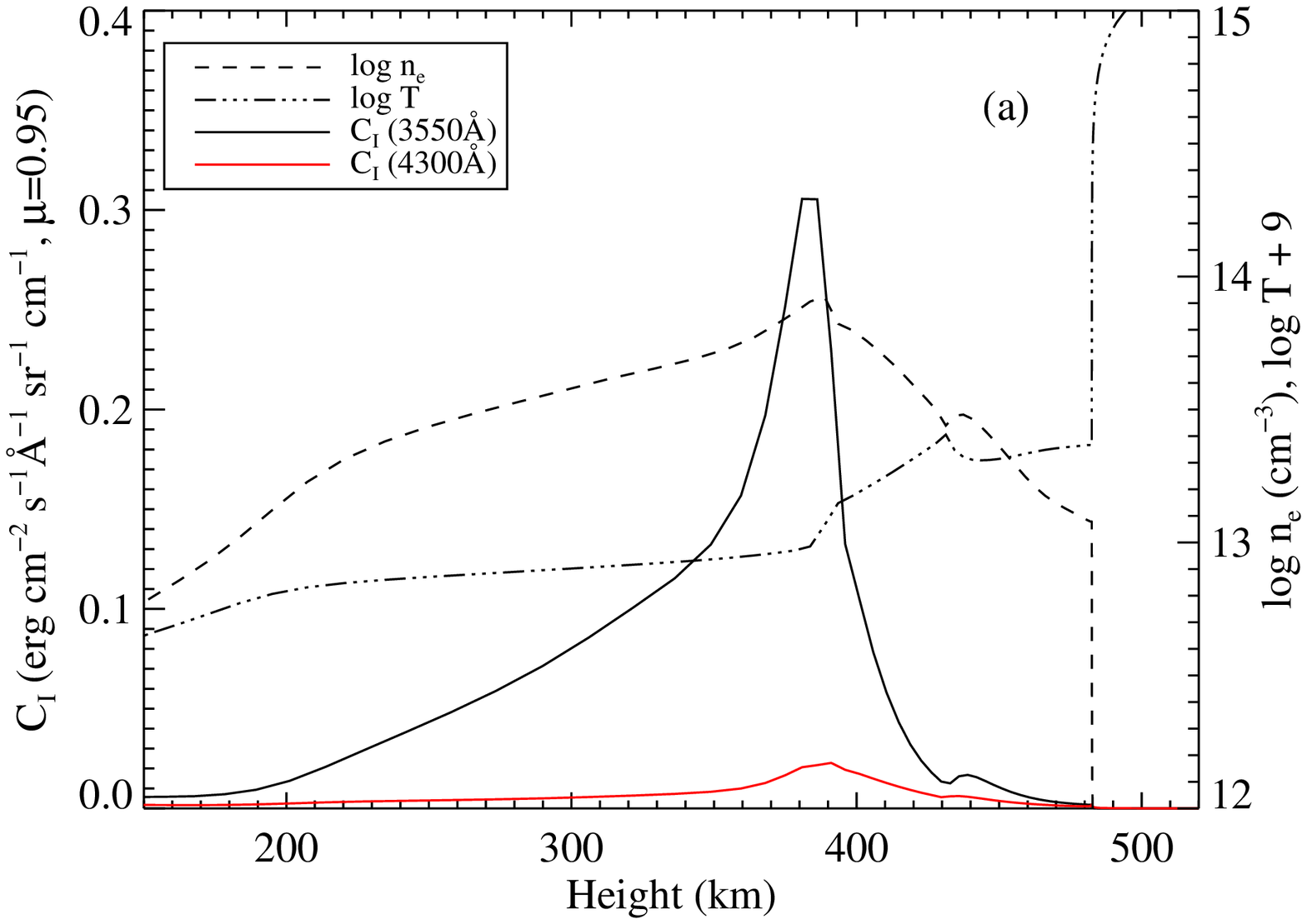}
\includegraphics[scale=0.4]{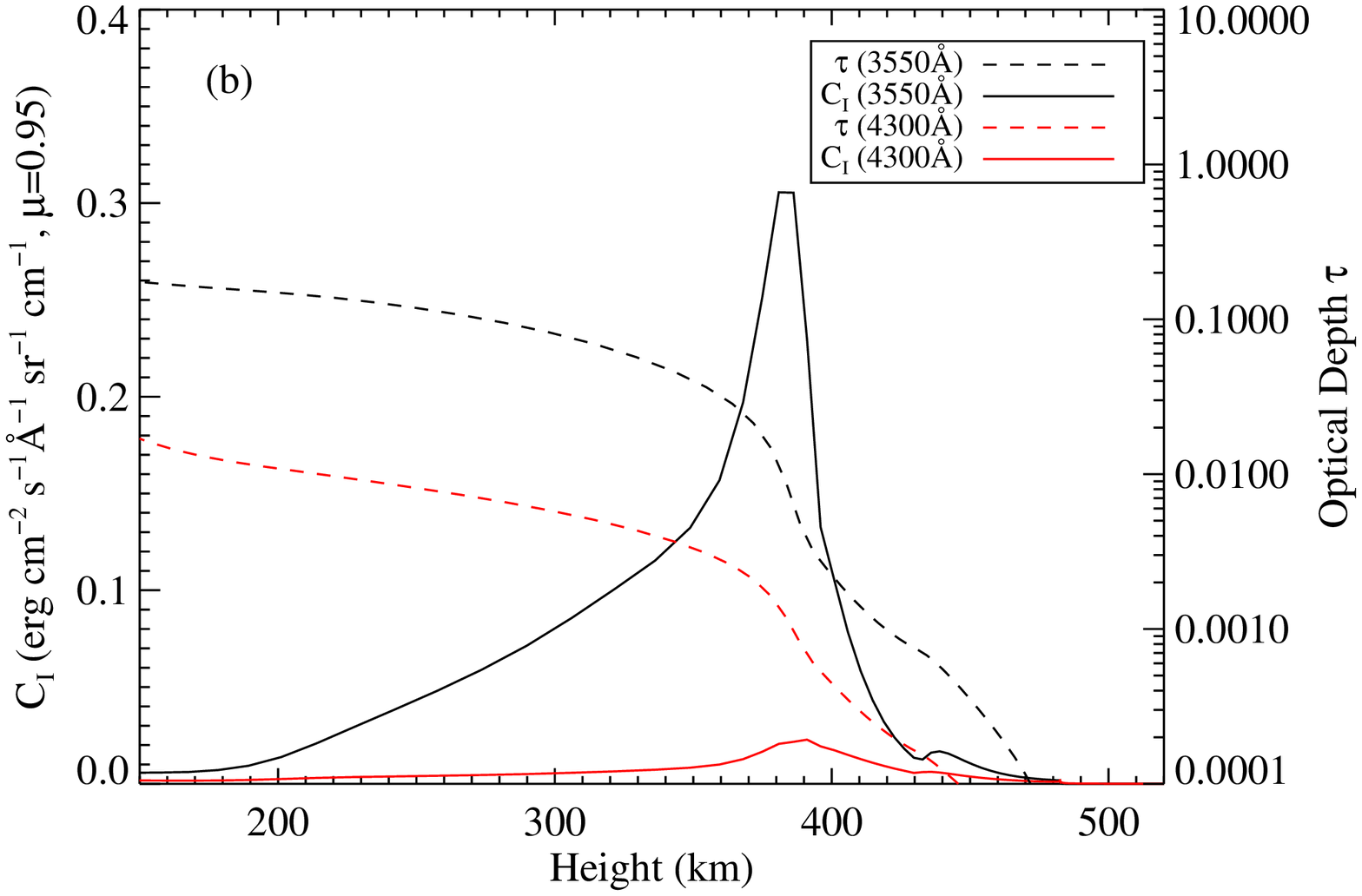}
\caption{ The contribution function for the F11 model at $\mu=0.95$ and $t=2.2$~s compared to the (a) electron density and temperature and 
 (b) optical depth.  Only heights corresponding to the flare chromosphere, transition region, and lower corona are shown.
}
\label{fig:contrib_f11}
\end{center}
\end{figure}

The contribution functions for the F11 model at $t=2.2$~s are shown in Figure \ref{fig:contrib_f11}(a\,--\,b). At this time,
the atmosphere has not yet experienced explosive evaporation, which is consistent with the F11 simulation from \cite{Allred2006}.
In contrast to the F13 simulation, the $C_I$ profiles for $\lambda=3550$\ \AA\ and 4300\ \AA\ during the F11 simulation extend over a large portion of the flare chromosphere.  The $C_I$ profiles also have second peaks near the preflare photosphere, which
is not shown here because the important molecular species and opacities in M-dwarf photospheres have not yet been included in RADYN (Section \ref{sec:method}).  The optical depths for 4300\ \AA\ and 3550\ \AA\ do not reach values of 1 until heights near the pre-flare photosphere;  
the values of $\tau_{3550}$ in the flare chromosphere range between 0.001 and 0.1, with  $\tau_{4300}$ about 
an order of magnitude less.  Therefore, the NUV and optical radiation from the flare chromosphere
  is dominated by optically thin hydrogen Balmer and Paschen continua emission from recombination due to the increased
electron density (and temperature).  The emergent F11 NUV and optical spectrum is a combination of mostly unattenuated 
 hydrogen recombination emission from $z=250-400$ km, combined with any increased photospheric radiation from backwarming \citep{Allred2006}.  
The value of $\chi_{\rm{flare}}$ is large ($\approx9$; Table \ref{table:observables}) and $T_{\rm{BB}}$ is low ($\approx5300$ K; Table \ref{table:observables}), as would be expected for hydrogen recombination at $T\approx10\,000 $ K with negligible wavelength dependent attenuation.

If we use the formula for spontaneous thermal hydrogen recombination emissivity \citep[Equation 7-2 of][see also \cite{Aller} and \cite{Wiese1972}]{Mihalas1978} in Equation (\ref{eq:didz}),
we obtain an approximate formula for the ratio of the contribution function at two wavelengths, $\lambda_1$ and $\lambda_2$:

\begin{equation} \label{eq:cont_ratio}
\frac{C_{I}(\lambda_1)}{C_{I}(\lambda_2)} \approx \frac{\lambda_2^2 n_2^3 g_{\rm{bf,1}}}{\lambda_1^2 n_1^3 g_{\rm{bf,2}}} e^{\frac{hc}{k_BT_e} (\frac{1}{\lambda_2}-\frac{1}{\lambda_1})} e^{1.578\times10^5/T_e (\frac{1}{n_1^2}-\frac{1}{n_2^2})} e^{(\tau_2-\tau_1)/\mu}
\end{equation}

\noindent at height $z$ with ($T_e, n_e$), where $g_{\rm{bf}}$ are the Gaunt factors and $n_1$ and $n_2$ are the principal quantum numbers
that dominate the opacity in Equation (\ref{eq:bf_ideal}) for $\lambda_1$ and $\lambda_2$, respectively\footnote{The value 1.578$\times10^5$ in the second
exponential is 13.598 eV / $k_B$.}.
When $\tau_{1,2}$ are negligible (as for the F11 in the flare chromosphere), 
$C_{I}(\lambda_1=4300)$/$C_{I}(\lambda_2=6690)$ for $\mu=1$
is $\approx0.8$ at $T=10\,000$ K and $e^{\tau_2-\tau_1}$ is not much larger than 1, which causes an optically thin Paschen
recombination continuum to exhibit a color temperature of $\approx5000$ K \citep[][; Section 9 of K13]{Kunkel1970, Kerr2014}.  
When $\tau_{1,2}$ are non-negligible (as for the F13 at $t=2.2$~s, Section \ref{sec:thick}), the ratio of the attenuation factors is $e^{\tau_2-\tau_1} > 1$ which causes the optical continuum to appear hotter.  Therefore, increasing the optical depth makes the Paschen continuum bluer with a higher value of 
$T_{\rm{BB}}$ (apparently hotter), as can be seen by approximating the atmosphere as a slab with constant ($T_e, n_e$) and integrating over height.
Note, the correction of the b-f opacity for stimulated emission (second term in Equation (\ref{eq:bf_ideal})) also affects the color temperature because it decreases the opacity at
 $\lambda=4300$\ \AA\ by a smaller factor than at $\lambda=6690$\ \AA.
However, the actual emergent intensity is an integration over height with varying ($T_e, n_e$), which is 
why we model the atmosphere gradients, which are apparent in Figure \ref{fig:dynamics}.  The emergent radiative flux is a Gaussian integration of intensity over five $\mu$ values:
the intensity at a smaller value of $\mu$ gives an even bluer optical continuum and smaller Balmer jump ratio
due to the $e^{(\tau_2-\tau_1)/\mu}$ factor.

\section{Stark Broadening at the Balmer Edge} \label{sec:nonideal}

Stark broadening is thought to be an important symmetric broadening mechanism of the hydrogen Balmer lines during solar and stellar flares \citep{Svestka1963,Worden1984,HawleyPettersen1991,JohnsKrull1997,Paulson2006,Allred2006}, in addition to possible turbulent \citep{Eason1992, Doyle1988} 
and thermal contributions.
 The linear (first order) Stark effect is the splitting of the degenerate orbital angular momentum ($l$) states of each principal
 quantum state ($n$) of hydrogen, 
 due to a net electric microfield from the surrounding distribution of charges.
 To date, the most complete theories of the linear Stark effect are those of 
 \cite{Vidal1970, Vidal1971, Vidal1973}, \cite{Seaton1990}, \cite{Stehle1993}, and \cite{Kepple1968},
which take into account the perturbations from the slowly moving (quasi-static) protons and from the 
 rapidly moving (dynamic) electrons.
 The Vidal \emph{et al.} treatment is known as the unified theory because it provides an accurate treatment of the electron 
perturbations from the line core through the far wings.  Theories of Stark broadening typically only present results for the lowest order 
 Balmer transitions, but \cite{Lemke1997} extended the results from the Vidal \emph{et al.} unified theory to $n=22$.  Therefore, this Vidal \emph{et al.}
 unified theory \citep[which is considered the best treatment of the non-overlapping far line wings and of the higher order Balmer lines; see the discussion in][ and in \cite{Vidal1970}]{JohnsKrull1997} is most useful to the interpretation of 
 flare spectra in the Balmer jump region\footnote{\cite{Bengtson1970} provided results up to H12 using the theory of \cite{Kepple1968} for one temperature and electron density.
 Bengtson (1996, private communication) provided higher order line profiles for the study of \cite{JohnsKrull1997}, but these results have not been made publicly available to our knowledge.}. 

In future work (Section \ref{sec:work}), we will implement the \cite{Lemke1997} extension of the Vidal \emph{et al.} results (see Section \ref{sec:work}).
In RADYN, the analytic approximations of the Stark damping from \cite{Sutton1978} are added to the total damping in the Voigt profile (Section \ref{sec:method}).  Solar modeling
of infrared hydrogen lines has shown that this is adequate \citep{Carlsson1992}. 

\subsection{Occupational Probability of Level $n$} \label{sec:totwn}
  Here, we address the atomic physics necessary to interpret the merging of the rapidly converging higher order Balmer lines and the 
 apparent lack of a Balmer discontinuity (or edge) at $\lambda=3646$\ \AA\ in flare spectra.

To model the Stark broadening of the Balmer edge, we apply the occupational probability 
theory developed in \cite{HM88} and \cite{Dappen1987}, which is summarized
thoroughly in the recent modeling study of cool DA white-dwarf atmospheres of \cite{Tremblay2009}.  This theory provides a
 thermodynamically self-consistent approach to pressure ionization, which allows the internal partition function to converge.
 Unlike procedures that truncate the internal partition function, occupational probability theory includes a 
continuous transition from bound to free states as upper levels
 of hydrogen converge and Stark energy shifts cause them to overlap.
Electric field ionization experiments with sodium have shown that enhanced ionization rates of electrons in level $n$ occur when there is a 
\emph{critical} electric microfield strength causing the highest Stark state of $n$ to cross with the lowest Stark state 
of $n+1$ \citep{Pillet1984}. At a level crossing, electrons can undergo a Landau-Zener transition from $n$ to $n+1$ as the 
electric field fluctuates with time above and then below the critical value \citep{Zener1932}. 
For electric field perturbations that cause level crossing of $n$ and $n+1$,
all states $\ge n$ become \emph{dissolved} or \emph{destroyed}, and an electron excited to $n$ can undergo a reverse-cascade to the continuum
through a series of Landau-Zener transitions. 

The occupational probability, $w_n$, is the probability of an atom being perturbed by a net electric
microfield that is below the critical field strength that would dissolve/destroy level $n$:  

\begin{equation} \label{eq:wn}
w_n (\rm{charged}) = \int_{0}^{\beta_{crit}} P(\beta) d\beta 
\end{equation}

\noindent where $P(\beta)$ is the probability distribution of net electric microfield strength ($\beta$) and 
$\beta_{\mathrm{crit}}$ is the critical net electric microfield that dissolves/destroys level $n$ \citep{Seaton1990}.
Both the high frequency electron collisions and quasi-static proton collisions contribute to level
dissolution, but $w_n$ is primarily determined from the ambient protons \citep{HM88, Tremblay2009}. 
The ambient electrons contribute to the cascade plus ionization process through their high-frequency collisions
with the electrons that are excited to the dissolved states and through the mixing of Stark states allowing 
the reverse-cascade of Landau-Zener transitions to proceed \citep[see][]{Tremblay2009}.  
Following \cite{Tremblay2009}, we adopt the \emph{Q-fit} electric microfield model \citep{Nayfonov}
of the Hooper distribution to represent the probability distribution $P(\beta)$.  In our calculations,
we use the \emph{total} occupational probability obtained 
by multiplying Equation (\ref{eq:wn}) by the probability of level destruction from neutral collisions, which are modeled using  
the same method described in \cite{Tremblay2009} and \cite{HM88}.  For the conditions in flare atmospheres,
the total occupational probability is approximately equal to $w_n(\rm{charged})$, given by Equation (\ref{eq:wn}).

\subsection{Continuous Opacities at the b-f Edge}
\cite{Dappen1987} developed the method for incorporating the occupational probability theory
into b-f and b-b opacities for hydrogen to account for the effects of Landau-Zener transitions and the 
subsequent cascade to the continuum. 

  The Landau-Zener b-b opacity for a Balmer line, H$n$,
is the nominal b-b opacity multiplied by the occupational probability $w_n/w_{n=2}$, where $w_{2}=$1 for the conditions in flare atmospheres.  The Landau-Zener b-f opacity is the nominal b-f opacity (Equation (\ref{eq:bf_ideal}))  
 extrapolated to wavelengths longer than the standard b-f edge ($\lambda_o$),
 multiplied by the probability that an atom experiences a critical microfield strength or greater,
 causing the upper level of the transition to be dissolved. 
For any wavelength $\lambda > \lambda_o$, this probability is obtained
 by using a ``pseudo'' quantum number $n^*$:

\begin{equation} \label{eq:nstar}
n^* = (\frac{1}{n_i^2} - \frac{hc}{\lambda Z^2 E})^{-1/2}
\end{equation}

\noindent where $E=13.598$ eV $=2.1787102 \times 10^{-11}$ erg and $Z$ is the atomic number ($Z=1$ for hydrogen).
The \emph{dissolved probability} (or \emph{dissolved fraction}) is 

\begin{equation} \label{eq:dissfrac}
 D(\lambda)=1-\frac{w_{n^*}}{w_{i}}
\end{equation}

\noindent where $w_{n^*}$ is the occupational probability for level $n^*$ and $i=2$ for the Balmer lines.  As discussed by \cite{Dappen1987} the
 pseudo levels ($n^*$) are simple extrapolation tools between the limits of $w_n \approx 1$ and $w_n << 1$ that are 
useful 
 for estimating the critical microfield strength (where the highest pseudo-stark state of $n^*$ and lowest of $n^*+1$ overlap)
that can cause level dissolution through a cascade of Landau-Zener transitions.  
Also, note, there are several different ways to estimate the critical field strength, and we use the popular method from \cite{Seaton1990}. 
 
\begin{figure}[H]
\begin{center}
\includegraphics[scale=0.6]{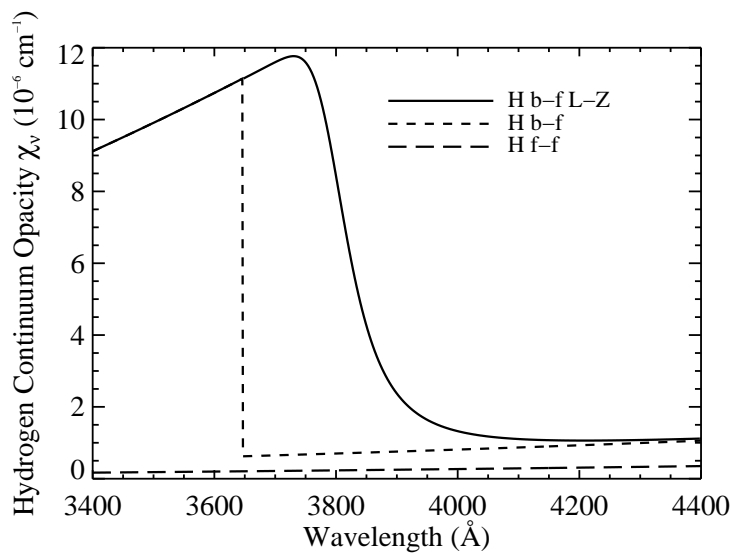}
\caption{ The hydrogen b-f continuum opacity accounting for Landau-Zener (L-Z) transitions, at $t=2.2$~s in the F13 simulation,
at a height of 255.4 km (log$_{10}$ $m =-2.57$ g cm$^{-2}$) which is the maximum of 
$C_{I}(\lambda=3550)\Delta z \times C_{I}(\lambda=4300) \Delta z$ where $\Delta z$ is the vertical extent of a grid cell.
  The values of $n_p$ and $T_e$ are respectively,
$5.6 \times 10^{15}$ cm$^{-3}$ and 12,770 K.  At this height, hydrogen is nearly 75\% ionized.
Also shown are the hydrogen f-f opacity and the 
 hydrogen b-f opacity without accounting for opacity effects from Landau-Zener transitions at this height.  
At this layer, $\tau_{3550}=1.1$ and $\tau_{4300}=0.2$.
}
\label{fig:wn1}
\end{center}
\end{figure}

Now accounting for Landau-Zener transitions, wavelengths longer than the Balmer edge ($\lambda_{\rm{o}}=3646$\ \AA) allow there to 
 be a fraction of atoms in $n=2$
that can be photoionized by $\lambda > 3646$\ \AA, whereas without Landau-Zener transitions, photons with $\lambda > 3646$\ \AA\ could
 photoionize atoms only with electrons in $n>2$.  The total b-f opacity, with and without Landau-Zener transitions,
 in a representative layer of the CC in the F13 model 
  is shown in Figure \ref{fig:wn1}.  Due to the high ambient proton density ($5.6\times10^{15}$ cm$^{-3}$), the dissolved fraction is 
  $>$99\% (the occupational probability, $w_n$, is $<1$\%) for the upper levels that contribute opacity at $\lambda \lesssim3730$\ \AA.
  The nearly complete level dissolution results in a direct extrapolation of the Balmer b-f opacity
 to these wavelengths.  As $\lambda$ increases, the fraction of $n=2$ hydrogen atoms that can be photoionized
decreases, due to the larger critical microfield required to cause destruction of the upper level of an atomic transition.
This leads to a continuous transition from the Balmer continuum opacity 
 into the Paschen continuum opacity, which dominates at $\lambda > 4200$\ \AA.  The hydrogen f-f opacity (without Landau-Zener transitions) is 
also shown in this figure, illustrating that it is small but non-negligible ($\chi_{\rm{f-f}}/\chi_{\rm{tot}}\approx0.2$) at redder wavelengths.

Hereafter, we refer to these new opacities as Landau-Zener opacities, and the pseudo-continuum resulting from this opacity
as the Landau-Zener continuum.  Note, \cite{Dappen1987} referred to this as the ``dissolved level pseudo-continuum.''
 In Figure \ref{fig:wn2}, we show how the Landau-Zener b-f opacity from $\lambda=3646$\ \AA\ to $\lambda=4200$\ \AA\ changes 
 as a function of $n_p$ ($\approx n_e$).  
  Self-consistent atmospheric parameters were obtained from representative heights in the F11 and F13 models, at various times, and 
covering a range of proton densities, $n_p = 10^{13} - 5\times10^{15}$ cm$^{-3}$. As the ambient charge density increases, 
the Landau-Zener b-f opacity maximum shifts from $\lambda=3660$\ \AA\ to $\lambda=3730$\ \AA.
 Also shown here are the occupational probabilities, $w_n/w_2$ (open circles, right axis, $w_2 = 1$),
that are multiplied by the nominal b-b opacity for each of the Balmer b-b transitions.  Thus the opacity from the lines is transferred to the 
Landau-Zener continuum.

\subsection{Landau-Zener Opacity Modeling Using the RH Code} \label{sec:wn}
We use the NLTE RH code \citep{Uitenbroek2001} to model the Landau-Zener b-f
 and b-b opacities and emissivities between $\lambda=3646$\ \AA\ and $\lambda=5000$\ \AA\ for the F13 model 
at $t=2.2$ s.
 The RH code is well-suited for modeling the Balmer jump region due 
to the ability to calculate overlapping transitions \citep{RH1992} with
 more wavelength points than in the original dynamic (RADYN) simulation. 
By assuming statistical equilibrium and hydrogen number conservation, we can use an arbitrarily large hydrogen atom to model the 
merging of higher order lines near the Balmer b-f edge. 
Statistical equilibrium is a poor assumption during the early phases of the flare simulation, but at $t=2.2$~s, the resulting error is small\footnote{We compared the RH and RADYN predictions at $t=2.2$~s for the Balmer lines and continua that overlap, and we found
 satisfactory agreement.}.  
 As in RADYN, the linear Stark broadening in RH is treated with a Voigt profile and the approximate formulae from \cite{Sutton1978} are used;
a prescription for quadratic Stark broadening is also included.  

We obtain the atmospheric parameters (column mass, temperature, electron density, and velocity) at $t=2.2$~s from the RADYN simulation with a six-level hydrogen atom,
and we calculate the spectrum with the RH code using a 20-level hydrogen atom.  
In the RH calculation, Ca \textsc{ii} and He are treated as background (LTE) elements.  
At each depth in the atmosphere, the effect of Landau-Zener transitions is
modeled by multiplying the Balmer b-f cross section by the dissolved fraction (Equation (\ref{eq:dissfrac})) and the Balmer b-b line profile by the occupational probability ($w_n/w_2$), as in 
\cite{Tremblay2009}.  The NLTE opacity and emissivity is then calculated in RH with these modifications.
We show the result of the RH flux calculation in Figure \ref{fig:pseudoC_S28}, 
compared to several observed flare spectra during the mid rise (S\#27)\footnote{The S\#'s refer to the 
time-sequential spectrum numbers over each flare analyzed in K13.}, late rise (S\#28), and peak phases (S\#31) of a large flare 
on the dM4.5e star YZ CMi from K13. The observed spectra have been normalized to the flux at $\lambda=4700$\ \AA\ to show 
the range of properties at $\lambda < 3900$\ \AA\ and the lack of color temperature evolution at $\lambda > 4000$\ \AA\ (see K13 for the evolution of the flare flux without scaling).  Because the flare area is unresolved, the continuum brightness is not constrained by observations.  Thus, the RH model surface-flux spectrum has been scaled by the factor $R_{\rm{flare}}^2/d^2$ (where $R_{\rm{flare}}$ is the projected radius of a circular flare area, and $d$ is the distance to the star) to match the $\lambda=4700$\ \AA\ flux of each observed flare spectrum.
  An instrumental convolution with a Gaussian of FWHM $=$ 13.5\ \AA\
has also been applied to the model.  We are not attempting to model this flare here, but we use these spectra
to illustrate that the model spectrum characteristics are similar to the data at least at some phases of the flare evolution.

 First, the blue-optical color temperature between the Balmer lines at $\lambda>4000$\ \AA\ 
is well fit to a blackbody with a temperature of $T_{\rm{BB}}=10\,400$ K, which is shown in Figure \ref{fig:pseudoC_S28} as the dashed line.  Note that this color temperature
is larger by 1000 K compared to that in Table \ref{table:observables}, due to the extension of the Landau-Zener continuum and Balmer wings between $\lambda=4000$\ \AA\ and $\lambda=4060$\ \AA
 \footnote{We use the same continuum blue window (BW) wavelength regions and weighting as in K13. }.
Although the color temperature is represented by a blackbody with $T=10\,400$ K, the F13 model radiative flux is larger by a factor of 1.7
compared to the surface flux of a blackbody with this temperature. The radiation temperature ($T_{\rm{rad, \mu=1,\lambda=4300}}\approx11\,600$ K) is also significantly larger than the color temperature ($T_{\rm{BB}}=9300-10\,400$ K).  Hereafter, we refer to the hot blackbody component as the hot ``blackbody-like'' 
continuum component, since
we have identified the emission as predominantly due to optically thick hydrogen (Paschen) recombination emission in Section \ref{sec:thick}.
Second, a jump in flux occurs above an extrapolation
of the blackbody to the bluest wavelengths, and this jump in flux is the same in both the F13 model and the observations of the late rise phase spectrum. 
 Third, there is no Balmer discontinuity feature
present in the model spectrum at $\lambda=3646$\ \AA:  the model calculation produces 
 a featureless Landau-Zener continuum connecting the Balmer continuum emission at $\lambda<3646$\ \AA\
to the bluest significant Balmer line in emission, H10 $\lambda3798$.  This effect originates from the direct extrapolation of 
Balmer continuum opacity in Figure \ref{fig:wn1} to these wavelengths where the opacity is due to the upper levels of hydrogen that have nearly
100\% probability of being dissolved.

The RH flux spectrum with the opacity effects from Landau-Zener transitions (and without instrumental convolution) 
is shown in Figure \ref{fig:pseudoC_S28_zoom} around the 
Balmer edge region; no Balmer lines 
are at all apparent between $\lambda = 3646 - 3740$\ \AA, and H11 is barely identifiable. 
The RH prediction using a 20-level hydrogen atom \emph{without} the effects from Landau-Zener transitions (gray line, ``no L-Z'') is shown and definitively illustrates 
that the overlap/merging of the Balmer wings indeed form a pseudo-continuum between the Balmer lines at wavelengths
shorter than 4100\ \AA, but less than the calculation with the opacity effects from Landau-Zener transitions (black line, ``L-Z'').  
The two calculations disagree most concerning the appearance and nature of the pseudo-continuum between 3646\AA\ and 3720\ \AA. In particular,
the ``no L-Z'' calculation exhibits a ``blue continuum bump'' at $\lambda \approx3720$\ \AA\ and a large positive slope at wavelengths between $3646$\ \AA\ and 3720\ \AA, whereas the ``L-Z''
 calculation predicts a continuum with the same slope as the observed Balmer continuum at all $\lambda<3720$\ \AA.
The ``no L-Z'' calculation also shows that 
the H10\,--\,H19 lines do not merge enough to conceal the Balmer discontinuity, which is just as apparent if we apply a large instrumental 
broadening.  
Using a hydrogen atom with more levels (we also tried a 35-level hydrogen atom) does not improve the modeling result:  the opacity
effects from Landau-Zener transitions are necessary to completely conceal the Balmer edge at these densities.
As discussed by \cite{Dappen1987}, the slope of the pseudo-continuum in the ``no L-Z'' calculation between 3646\AA\ and the blue continuum bump 
is the result of the rapidly decreasing oscillator
 strength \emph{density} of the higher-order transitions.  In the ``L-Z'' calculation, Stark broadening decreases the oscillator strength density
 of the b-b transitions while replacing oscillator strength density with the Landau-Zener continuum.

In Figure \ref{fig:contrib_cont_3870}, we show how the contribution function at $\lambda=3870$\ \AA\ (between H8 $\lambda3889$ and H9 $\lambda3835$) and $\lambda=4300$\ \AA\ compare with and without opacity effects from Landau-Zener transitions.  As expected, the contribution at $\lambda=4300$\ \AA\ does not change.  
For $\lambda=3870$\ \AA\ without Landau-Zener transitions, 
the opacity is dominated by Paschen b-f opacity. This wavelength is more optically thin than $\lambda=4300$\ \AA.  Hence there is more contribution from the stationary component of the flare atmosphere at $z=150-230$ km.  In the case with Landau-Zener transitions,
the Landau-Zener opacity increases the optical depth beyond that of the optical depth at 4300\ \AA\ 
(the contribution at $z=150-230$ km is less than in the calculation without Landau-Zener transitions) 
and more $\lambda=3870$\ \AA\ photons
escape from higher in the atmosphere in the CC (at $z=251$ km) due to recombination of electrons from the $n^* = 8.3$ pseudo-level (Equation (\ref{eq:nstar})) to the $n=2$ level of hydrogen.  

We can now associate the model flux spectrum in Figure \ref{fig:pseudoC_S28} to several atomic emission processes. 
 At $\lambda < 3646$\ \AA\ the flux is almost entirely Balmer continuum emission from free electrons recombining to $n=2$.  
This part of the spectrum also exhibits a color temperature $T_{\rm{BB}}>10\,000$ K, with a peak in the NUV between $\lambda=2500-2600$\ \AA.
The Landau-Zener continuum flux between $\lambda=3646-3760$\ \AA\ is \emph{bona-fide}
Balmer continuum recombination radiation resulting from free electrons recombining from the $n^*$ pseudo (dissolved) upper levels of hydrogen to $n=2$ \citep{Dappen1987}; \emph{i.e.}, undergoing reverse Landau-Zener transitions from the continuum to $n=n^*$, at which point they transition to $n=2$.
The flux between the rapidly converging Balmer lines from H10 to H$\epsilon$ is mostly 
 Landau-Zener continuum emission from electrons recombining from the dissolved upper levels to $n=2$, but at these wavelengths
there is a smaller fraction of atoms in the atmosphere that experience a critical field strength, and thus the Landau-Zener continuum emission is less than at $\lambda=3646-3760$\ \AA.  There is also some contribution from the far wings of the Balmer b-b transitions.  
The far wings of H$\delta$ and H$\epsilon$ and the redmost extent of the Landau-Zener continuum emission 
produce additional emission above the nominal Paschen recombination spectrum as red as $\lambda=4200$\ \AA\ (compare solid and dotted red lines).  
At $\lambda>4200$\ \AA, Paschen continuum emission from recombination of free electrons to $n=3$ dominates, with a smaller ($\approx$20\%) contribution 
from thermal bremsstrahlung (free-free) emission.  Paschen recombination radiation also contributes toward the radiation at $\lambda < 4200$\ \AA\
but to a much smaller degree than the calculation without including the Landau-Zener continuum, as can be seen by the attenuation of radiation at 
$z=160-230$ km in Figure \ref{fig:contrib_cont_3870}.  An interesting observational effect of including the Landau-Zener continuum emission is that the measured color temperature of the flux spectrum increases from 9300 K (6 level, without Landau-Zener transitions, dotted red line, Table \ref{table:observables}) to 10\,400 K (20 level, with Landau-Zener transitions, thick red line), improving further the applicability of the F13 model to the measured $T_{\rm{BB}}$ distribution at peak for a variety of flares (K13).

We note that the lower order Balmer lines (H$\alpha$-H$\delta$) can be used 
to constrain the models.  The CC in the F13 model generates broad profiles, large surface fluxes, and a reverse decrement (more H$\gamma$ flux compared to H$\alpha$ flux) in the lower order Balmer lines.  
 The large widths of the lower order Balmer lines in Figure \ref{fig:pseudoC_S28}
 result from Stark broadening. However,
improvements to the implementation of the Stark effect (see Section \ref{sec:work}) are required for a detailed comparison.
Although the F13 model produces approximately $18$ times the surface flux in H$\gamma$ compared to the F11 model (and approximately $6$ times that in the F12 model), the ratio of Balmer line to continuum values is notably lower than 
in the smaller NT electron flux simulations (Table \ref{table:observables}).  At $t=2.2$~s, 
the relative amount of Balmer line to continuum emission produced by the 
CC is consistent with typical observed values in the impulsive phase:  H$\gamma_{\rm{EW4170}} \approx 20$ in the F13 model compared to
H$\gamma_{\rm{EW4170}} \approx 15-30$ in the late rise and peak phase spectra in Figure \ref{fig:pseudoC_S28}. 
Additional flaring areas that are either impulsively heated with smaller NT electron flux or gradually decaying from a large electron flux may be required to more accurately reproduce the time evolution of the emission lines in these individual spectra.

\subsection{A New Diagnostic of Ambient Charge Density in Chromospheric Flares} \label{sec:wn2}
The opacity effects from Landau-Zener transitions 
 result in Balmer recombination radiation longward of 3646\AA\ and explain the NUV/blue ``pseudo-continuum'' previously attributed to the 
merging of Stark-broadened Balmer lines.  The 
occupational probability theory as employed in white-dwarf atmospheric models provides a self-consistent quantum
mechanical method of modeling the opacity effects from cascading Landau-Zener transitions.  
This new modeling explains the lack of an observed Balmer discontinuity (or edge) in flare spectra,
while still allowing a \emph{Balmer jump} in flux above an extrapolation of a blackbody that is fit to 
the blue-optical wavelengths in the Paschen continuum ($\lambda > 4000$\ \AA).
Accounting for Landau-Zener transitions gives an emergent flux spectrum from the F13 model at $t=2.2$~s with continuum emission 
bridging Balmer continuum emission at $\lambda \le 3646$\ \AA\ to the highest-order Balmer line that is detectable, which is H10 or H11 
in the case of the F13 model.
The microscopic picture that is reinforced here is that 
slowly moving thermal electrons are able to recombine to $n=2$ and produce a photon with wavelength $>$3646\AA\ (a Balmer continuum
photon) due to the nearly complete dissolution of the upper levels: $1-w_n/w_2 \ge$0.85 for levels $n\ge11$, in a dense plasma with $n_p=5\times10^{15}$ cm$^{-3}$.
The success of modeling the ``pseudo-continuum'' emission from $\lambda=3646-3760$\ \AA\ in Figure \ref{fig:pseudoC_S28} as a red wavelength
extension of Balmer continuum emission 
is consistent with our conclusion that the white-light emission, in this RHD model, is dominated by a combination of hydrogen Balmer b-f, hydrogen Paschen b-f, and hydrogen f-f emission.

In solar flares, a ``pseudo-continuum'' component between the Balmer lines from $\approx3700-3870$\ \AA\ was first noted by \cite{Zirin1980} and \cite{Zirin1981}.  
The current, best treatment of the modeling of the Balmer jump region and pseudo-continuum
 was applied to several solar flares by \cite{Donati1985},
who used static, isothermal models with $n_e=10^{13.2}-10^{13.6}$ cm$^{-3}$ and $T_e=7000-10,000$ K.  
These models produced a ``blue continuum bump'' at $\lambda=3675$\ \AA, which shifts   
to redder wavelengths and increases in brightness for higher electron density, and has been discussed in light of a new 
solar flare observation with broad wavelength coverage spectra in the NUV \citep{Kowalski2015}. 
The pseudo-continuum bump at $\lambda \approx 3720$\ \AA\ in Figure \ref{fig:pseudoC_S28_zoom} produced in our 
 calculation without Landau-Zener opacity effects (Section \ref{sec:wn}) is very 
likely analogous to the NUV/blue bump feature at $\lambda \approx 3675$\ \AA\ produced in the \cite{Donati1985} lower density slab models.
As noted by these authors, this feature is due to the merging of Stark broadened high order Balmer lines.
Another feature in the \cite{Donati1985} models is a continuum ``dip'' at 3646\AA\ due to the smearing of the Balmer edge.
 \cite{Donati1985} truncated the model hydrogen atom at $n=84$ and broadened the Balmer edge using the Stark width of the highest 
Balmer line.  For $n=84$, the Stark damping parameter is 30\ \AA\ \citep{Svestka1963, JohnsKrull1997} which results in a flattening and smearing of the 
edge in their models.
Since the Landau-Zener continuum modeling from occupational probability theory does not predict 
a continuum bump at $\lambda\approx3720$\ \AA\ or a dip near the Balmer edge (for very high proton densities), 
this modeling may also provide a more self-consistent
explanation for the range in Balmer jump characteristics observed in solar flare spectra \citep[as in the compilation by][]{Neidig1983}, while
also improving the diagnostic capabilities of these spectra.

Intuitively, occupational probability theory provides insight beyond a model that considers pseudo-continuum from merged lines as just a superposition  
of b-b opacities.  In the classical framework \citep{Donati1985}, 
the opacity of a b-b transition is spread over a larger spectral range by the Stark broadening, and is then added to
the opacity of other Stark-broadened b-b opacities at a given wavelength.  In the occupational probability framework, the Stark-broadened b-b opacity is 
replaced by a larger b-f opacity at a given wavelength when hydrogen atoms experience a critical microfield perturbation. The larger b-f opacity results from the fact that free states are not degenerate (in orbital angular momentum), and therefore the oscillator strength density of the 
continuum is not re-distributed over a superposition of Stark states as for b-b transitions.  This physical difference
explains the large discrepancy between the two model spectra in Figure  \ref{fig:pseudoC_S28_zoom} from $\lambda=3646-3730$\ \AA\ \citep[see also the discussion in][]{Dappen1987}.  In summary, the opacity sources from \cite{Donati1985} are \emph{continuum opacity from a Stark-broadened b-f edge and a superposition of Stark-broadened b-b opacities} whereas in occupational probability theory,
the opacity sources are \emph{continuum opacity from Stark-dissolved levels and a superposition of the remaining undissolved Stark-broadened b-b opacities.}

\begin{figure}[H]
\begin{center}
\includegraphics[scale=0.6]{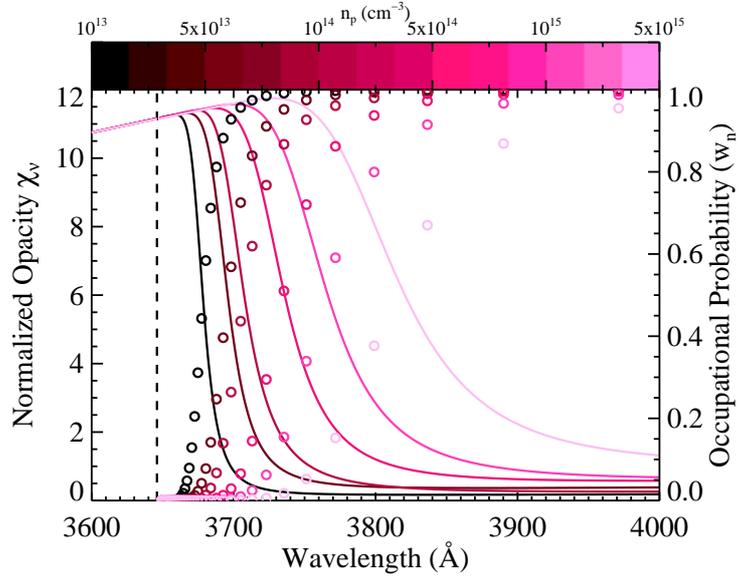}
\caption{ 
The hydrogen b-f opacity accounting for Landau-Zener transitions at representative times and heights in the F11 and F13 simulations. Each curve corresponds
to approximately the ambient charge density at each of the tick marks in the colorbar.  The opacity curves have been
scaled to a common value at $\lambda=3600$\ \AA\ to illustrate the wavelength shift of the opacity maximum 
compared to the Balmer b-f edge (vertical dashed line) without accounting for Landau-Zener transitions.  The occupational probabilities (right axis) of the upper levels of the Balmer transitions are shown as circles (for reference H10 is at $\lambda \approx 3800$\ \AA).
}
\label{fig:wn2}
\end{center}
\end{figure}

\begin{figure}[H]
\begin{center}
\includegraphics[scale=0.6]{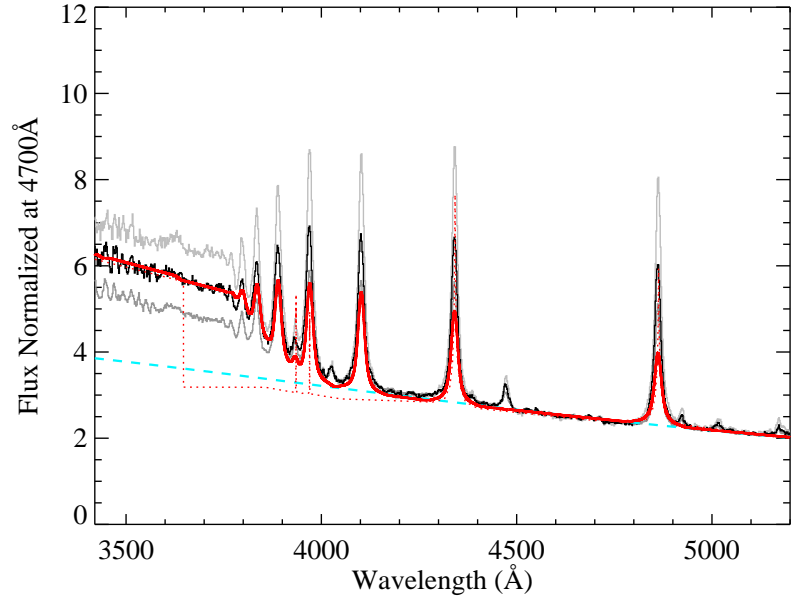}
\caption{ RH calculation with a 20-level hydrogen atom and the b-f and b-b opacities including Landau-Zener transitions, 
during a snapshot of the F13 simulation at $t=2.2$~s (solid red).   A $T=10\,400$ K blackbody fit to continuum windows at $\lambda=4000-4800$\ \AA\ is also shown (dashed light blue), along with the RADYN prediction with a 6-level hydrogen atom (red dotted line).  Instrumental broadening has been applied to the red spectrum.  Several observations during the rise and peak phases of the IF3 flare of K13 are shown: peak spectrum is S\#31 from K13 (dark gray), mid rise phase is S\#27 from K13 (light gray), and late rise phase is S\#28 from K13 (black).  All spectra are scaled to the flux at $\lambda=4690-4710$\ \AA.
}
\label{fig:pseudoC_S28}
\end{center}
\end{figure}

Modeling the opacity effects from Landau-Zener transitions provides a new physical interpretation of several dMe flare spectra in the literature.  
We've already shown how this modeling with a very large density ($\approx5\times10^{15}$ cm$^{-3}$) can account for the continuum and line properties in a flare with H10 or H11 as the bluest identifiable Balmer line (Figure \ref{fig:pseudoC_S28}).  In the continuum peak (1:15 UT) spectrum from the dMe flare studied in \cite{GarciaAlvarez2002}, H9 or H10 also appear as the bluest Balmer line; however,
their gasdynamic modeling using the \cite{Jev1998} fit to the Balmer decrement indicates densities nearly two orders of magnitude lower than the occupational probability modeling indicates based on the bluest identifiable Balmer line.  
 The bluest identifiable Balmer lines in most dMe flares with moderate spectral resolution are typically H15 and H16 both in the 
impulsive phase \citep{HawleyPettersen1991} and in the gradual decay phase \citep{Kowalski2010, Kowalski2013}.
Even $R=40\,000$ echelle spectra of dMe flares sometimes do not show emission in Balmer lines bluer than about H15 and 
instead exhibit a nearly featureless\footnote{Aside from many faint metallic and helium emission lines.} 
continuum with a color temperature of $9000-11\,000$ K at $\lambda \lesssim 3700$\ \AA\ \citep[Figures 2 and 5 of ][]{Fuhrmeister2008}.
It is also notable that one peak flare spectrum of a dMe flare showed Balmer lines as blue as H19 \citep{Fuhrmeister2011}, which the authors concluded 
was a result of only a small enhancement of the chromospheric electron density in the flare.
Since the Landau-Zener continuum resulting from occupational probability theory can fully account for similar characteristics of this 
 spectral region in the flare spectra
in Figure \ref{fig:pseudoC_S28}, we suggest the \emph{range} of characteristics observed around the Balmer jump in stellar 
flares may be explained with the range of Landau-Zener opacities in Figure \ref{fig:wn2}.  

Future modeling of particular flares using occupational probability theory can be used to infer the magnitude and gradient of the ambient charge (proton) density in
 the densest regions of the flare atmosphere that are significantly heated and produce white-light emission.
In Section \ref{sec:wn} and Figure \ref{fig:wn2}, we showed how the Landau-Zener continuum opacity maximum changes and occupational probabilities dissolve the hydrogen Balmer transitions over 2.5 orders of magnitude in charge density.  
 We present a simple method using the Balmer decrement to estimate the charge density for an optically thick flare atmosphere in LTE 
 (as for the F13 atmosphere studied here).  
We introduce the Landau-Zener Balmer decrement, which is the flux in a Balmer line transition (\emph{e.g.}, H10 or H11 which are relatively unaffected by strong metallic or helium lines) that experiences a large change in the dissolved fraction of the upper level of this transition,
 divided by the flux of a lower order line that is not affected by level dissolution (\emph{e.g.}, H$\gamma$)\footnote{Assuming that the Balmer line emission does not originate at a different spatial location, for example from a larger area with a lower density, than the continuum emission.}.  
In Table \ref{table:wndec}, we show how the Landau-Zener Balmer decrement H11/H$\gamma$ changes for different temperatures and over the density range of $n_p=10^{13}$ to $n_p=5\times10^{15}$ cm$^{-3}$.  This Balmer decrement
is obtained by multiplying $w_{n=11}/w_{n=5}$ (where $w_{5}\approx 1$) by the Balmer decrement formula for the optically thick, isothermal, LTE limit in Appendix B of \cite{Drake1980}.  The decrement sharply
drops for $n_p > 10^{15}$ cm$^{-3}$ as $n=11$ becomes dissolved (see also Figure \ref{fig:wn2}).
The observed decrements for H11/H$\gamma$ in the two peak spectra from \cite{HawleyPettersen1991} are 0.26 and 0.22, which suggests a charge density of not much larger than $10^{15}$ cm$^{-3}$ but also not smaller than $10^{14}$ cm$^{-3}$.  This method could be used as an alternative to the Inglis-Teller relation\footnote{See the discussion in \cite{HM88} about how the theory of occupational probability differs from the Inglis-Teller relationship:  the Landau-Zener transitions
require \emph{fluctuations} of the microfield around the critical field value. }, which requires data with high spectral resolution.  
However, since lines and continua are formed at different physical depths and depth ranges (as for the F13 atmosphere studied here), we plan to improve the accuracy of this spectral diagnostic using a grid of NLTE flare atmospheres 
in future works.

\section{The Viability of the F13 Model} \label{sec:real}
M-dwarf flares are more energetic than solar flares (in total).  Typical white-light energies of dMe flares are $10^{30} - 10^{32}$ erg \citep{Hawley2014} with occasional events exceeding $10^{34}$ erg in the $U$-band alone \citep{Kowalski2010}.  
The areal extent of flare footpoints cannot be measured directly\footnote{Emission model fits, \emph{e.g.}, with a blackbody function or RHD model spectra, give an indirect estimate of the flare area and energy flux requirements.}, and hard X-ray emission is far too faint 
except during the largest dMe flares \citep{Osten2010}.  Although we cannot determine if NT electron fluxes on M-dwarfs are indeed as high as 
$10^{13}$ erg cm$^{-2}$ s$^{-1}$, typical values of $\approx10^{11}$ erg cm$^{-2}$ s$^{-1}$ in strong solar flares
 cannot explain the properties of white-light emission during dMe flares, and thus we infer that the energy flux is higher.  
The atmospheric response to an F13 NT electron burst provides insight into the radiative transfer 
processes that could explain white-light emission, but is this very high flux a realistic model for dMe flares?  
In this section, we
evaluate its viability and conclude that some issues still remain.

First, is an F13 beam even possible during an M-dwarf flare? 
The energy flux available to accelerate particles at the reconnection site is limited by the free magnetic energy in the corona, which can be estimated
by $\frac{B_{\rm{cor}}^2}{8\pi} v_{\mathrm{inflow}} = 2 \times
10^{13}$ erg cm$^{-2}$ s$^{-1}$, where
$v_{\mathrm{inflow}}$ is the rate at which magnetic field flows into
the reconnection region and may be between
$0.01-0.1$ times the Alfv\'en velocity. The energy flux requirement of $2\times10^{13}$ erg
cm$^{-2}$ s$^{-1}$  into each magnetic footpoint
assumes equipartition of energy among protons and electrons and that
all free magnetic energy goes into accelerating particles.  
Using the larger estimate for inflow velocity, $0.1 \times
v_{\rm{Alfven}}$ gives $B_{\rm{cor}}\approx 1.5$ kG as the coronal
magnetic field (assuming $v_{\rm{inflow}}=0.01v_{\rm{Alfven}}$ gives 3.3 kG as
the coronal magnetic field) necessary to supply this energy flux.  
Direct (Zeeman broadening) measurements of small-scale magnetic fields of M-dwarfs reveal average photospheric
magnetic flux density of $3-4.5$ kG \citep{JKV1996, PhanBao2009}, but the coronal
magnetic field above these starspots has not been directly measured.  The most well-studied active corona is that of the dM3.5e star EV Lac \citep{Osten2006},
which exhibits a very hot (10 MK), confined ($L\approx10^8$ cm), and dense ($n_e \approx 10^{13}$ cm$^{-3}$) component in quiescence.  
For these structures, the minimum magnetic field required for confinement is estimated to be $\approx$1 kG.  Other values from microwave data of flares and quiescence
have suggested coronal magnetic fields on M-dwarfs exceeding 1 kG
\citep{Osten2006, Smith2005}, which have also been inferred from
magnetic confinement estimates using the Haisch Simplified Analysis of
 data from the \textit{Extreme Ultraviolet Explorer} (EUVE) of many M-dwarf flares \citep{Mullan2006}.   Thus, the energy requirement for accelerating an F13 NT electron flux requires relatively high (kilo-Gauss) coronal magnetic fields but not outside the realm of current measurements.

Although the energy flux requirements may be plausible (given large M-dwarf coronal magnetic fields), 
the electron number flux for the F13 model may be problematic due to return-current effects.  
The electron number flux corresponding
to our double power law F13 spectrum is $10^{20}$ electrons cm$^{-2}$ s$^{-1}$.  
A return-current electric field is established from the charge displacement of electrons and protons, 
given that both particles are energized equally in the process of magnetic reconnection and retraction. 
The initial energy lost \emph{per} NT electron from the return-current electric field is 13 keV Mm$^{-1}$ at the apex of our
model loop \citep[Equation 29 of][]{Holman2012}.  The NT electrons with the cutoff energy (37 keV) lose all of their energy and are thermalized
after penetrating only 3 Mm into the atmosphere, after which the flux
of ongoing NT electrons decreases with distance as more NT electrons
are thermalized.  \cite{Holman2012} derives the evolution of the NT electron flux distribution accounting for losses by the return-current electric field.  We use these results and find that 
only NT electrons with $E_{\rm{initial}} > 85$ keV penetrate to the top of the chromosphere without thermalizing.
Furthermore, the F13 NT electron density is comparable to the ambient 
 density ($n_e=2.6\times10^{10}$ cm$^{-3}$) at the apex of our model coronal loop; this condition would cause the ambient
electrons to be accelerated and would result in instabilities and other effects that are not modeled here.

A more basic issue is that the number flux of electrons available in the
corona flowing into the reconnection site is only $\approx5\times10^{18}$ electrons
cm$^{-2}$ s$^{-1}$ (using $n_e \times 0.1 v_{\rm{Alfven}}$, where $B=1.5$ kG is obtained from the energy flux), whereas the number flux in the F13 beam is $10^{20}$ electrons cm$^{-2}$ s$^{-1}$. Even if electrons in the coronal loop
were globally accelerated, there are only an additional $4\times 10^{19}$ electrons cm$^{-2}$ s$^{-1}$ available.
%This problem is similar to the number problem in solar flares discussed in \cite{Fletcher2008}.
However, this simple assessment does not account for  processes that may increase electron 
density during reconnection and retraction \citep[\emph{e.g.},][]{Longcope2011}.  Furthermore, high densities between $10^{12}-10^{13}$ cm$^{-3}$ 
have been inferred for some (non-flaring) coronal structures on the Sun \citep{Feldman1994} and on the flare star EV Lac \citep{Osten2006}.
In summary, the energy and number flux requirements for an F13 beam could be met under certain conditions (a dense corona with high magnetic field), but the return-current losses
would be significant.

A final issue with the F13 model is the short presence of the hot, blackbody-like emission, which
appears only at the end of the impulsive phase between $t\approx1.8$~s and $t=2.3$~s. 
We expect a coadded burst spectrum to be more directly comparable to spectral observations, which
have integration times of 10~s or more. 
The coadded spectrum over the 5.1~s duration of the impulsive heating 
and the decay phases (dot-dashed light blue line in Figure
\ref{fig:cont_evol}) exhibits a 
lower color temperature (7500 K) and larger Balmer jump ratio (2.7)
than the instantaneous model spectrum
% filling factor of 1.93, very close to a blackbody.
at $t=2.2$~s (Table \ref{table:observables}).  
The coadded F13 spectrum is not as representative of
the peak phases of the M-dwarf flares as the instantaneous $t=2.2$~s
spectrum, which exhibits prominent $T_{\rm{BB}}\approx10^4$ K blackbody-like
emission as observed.  A longer duration of NT electron
energy deposition for $5-10$~s may produce similar coadded and instantaneous spectra,
but the simulation becomes very computationally intensive past 2.3~s with sustained heating (Section \ref{sec:dynamics}).

The coadded 5~s F13 spectrum is more representative of the atmospheric conditions in  
the gradual decay phase and the very early rise phase.  The gradual
decay phase of flares exhibits a Balmer jump ratio of $2.6-3.3$ and a
blue-optical color temperature of about 8000 K (K13), which are
similar to the observables obtained from the coadded F13 spectrum (Table
\ref{table:observables}).  The coadded F13 spectrum is also consistent with a spectral observation from 
the very early rise phase of a flare from K13, which we show in 
 Figure \ref{fig:coadd}.  K13 characterized two additional properties of flare spectra that  
 can constrain models of the time-evolution: 
the slope in the NUV at $\lambda < 3646$\ \AA\ and the amount of deviation
at red wavelengths (at $\lambda > 5000$\ \AA) from a blackbody fit 
to the blue-optical range at $\lambda=4000-4800$\ \AA.  Although observations of both the early rise phase and gradual decay
phase exhibit a Balmer jump ratio that are larger than during the peak phase, only the early rise 
exhibits a $T\approx$10\,000 K color temperature at $\lambda < 3646$\ \AA\
and a small deviation from the blackbody at redder wavelengths. These two common properties of the average
burst model and the early rise phase observation can be seen in Figure \ref{fig:coadd}.
However, we expect that 
flare emission at any particular time (even in the very early rise phase) may consist of a 
superposition of many individual bursts.  In order to conclude that any individual model can explain the time-evolution 
at any particular time, 
future modeling work should incorporate a 
treatment of the spatial evolution of flares, such as the development of two
ribbon arcades and sequentially heated white-light
kernels.

\section{Summary and Conclusions} \label{sec:discussion}

We have simulated the dynamic and radiative response of an M-dwarf atmosphere 
subject to a short burst of a constant, high flux (10$^{13}$ erg cm$^{-2}$ s$^{-1}$) of NT electrons with a moderately high
low-energy cutoff (37 keV).  
This simulation produces a heated ($T=12\,000-13\,500$ K), high density
($n_{H} > 10^{15}$ cm$^{-3}$) ``chromospheric
condensation'' with $N_{H}(n=2)>3\times10^{17}$ cm$^{-2}$, $N_{H}(n=3)>10^{17}$ cm$^{-2}$,
and a total physical depth range of $\approx 20$ km, resulting in large hydrogen b-f and f-f opacities within a short amount
of time (2.2~s) after the start of non-thermal electron heating.  At this time, the
``flare photosphere'' ($\tau_{5000}=1$) is located in the CC at a height that is nearly 250 km higher than the pre-flare
photosphere, and the density of the CC is over an order of magnitude larger than the material at this height in the pre-flare.
  Significant heating to $T>10\,000$ K at high densities (comparable
to the density in the very low chromosphere of the pre-flare) is achieved and
maintained in the CC and also in the stationary, slightly less dense regions of the
atmosphere $\approx$50 km below the CC (Figures \ref{fig:dynamics},
\ref{fig:didz1}).  

The atmospheric response in the F13 model provides a new, accurate
description of the radiation field in a flare atmosphere with a dense CC, and
moreover, shows that 
such an atmosphere can produce bright white-light emission with 
properties that are in good agreement with the observations. 
The flux spectrum from the F13 model at $t=2.2$~s exhibits a blue-optical color temperature of $T_{\rm{BB}}\approx10^4$ K between $\lambda=4000-4800$\ \AA\
 and a relatively small Balmer jump ratio ($\chi_{\rm{flare}} \approx2.0$;
 Table \ref{table:observables}), both of 
which are large improvements compared to the range of these parameters in lower beam flux models (F10, F11, and F12 keeping all 
else the same with the NT electron spectrum).  
These properties describe typical peak spectra of dMe flares, and we showed the
striking match of the instantaneous F13 model at $t=2.2$~s to the observations of one large dMe flare with time-resolved spectra in the impulsive phase (Figure \ref{fig:pseudoC_S28}).
The F13 model is the first RHD model to reproduce three generally observed characteristics 
($T_{\rm{BB}}$, $\chi_{\rm{flare}}$, and H$\gamma_{\rm{EW4170}}$; Table \ref{table:observables}) 
self-consistently. 

 The pioneering gasdynamic simulations of
\cite{Livshits1981} and \cite{Katsova1997} produced a CC with similar characteristics to
those in the F13 CC, and also $\Delta \tau \approx 1$ at $\lambda= 4500$\ \AA\
and $T\approx9000$ K, albeit with $10-30$ times 
lower NT electron energy fluxes.  It is interesting that at least an order of magnitude difference in heating rates gives 
similar dynamic responses in gasdynamic modeling and radiative-hydrodynamic modeling. 
 Our new radiative-hydrodynamic models use a more realistic
starting dMe atmosphere with a NLTE treatment of the entire atmosphere 
(photosphere, chromosphere, transition region, and corona), a detailed
treatment of the helium level populations (which are important for regulating
the onset of explosive evaporation; see Section \ref{sec:dynamics}), and a modern parameterization of the NT electron 
spectrum from RHESSI data.  However, a higher cutoff of the NT electron spectrum (as employed in our model) is expected 
 to require a larger flux in order to produce similar dynamics \citep{Fisher1989}.  Finally, 
the RADYN and RH codes provide a detailed
spectrum, which we have shown is important for critically testing the models 
against the observations, for revealing new atomic physics in flares
around the Balmer jump, for constraining the charge density in the
densest atmospheric regions that are heated during flares (Sections \ref{sec:wn}, \ref{sec:wn2}),
and for understanding how hydrogen recombination appears as a  
blackbody-like spectrum with $T\approx10^4$ K (Section \ref{sec:opaccontrib}).

By analyzing the contribution function for the continuum intensity, we
conclude that the values of the slope ($T_{\rm{BB}}$) and Balmer jump ratio
($\chi_{\rm{flare}}$) of the emergent optical and NUV continuum intensity
in the F13 flare atmosphere are due primarily to hydrogen recombination radiation subject to
wavelength dependent attenuation from Balmer and Paschen b-f
opacities.  
Laboratory spectra of hydrogen at high density ($n \approx 10^{16}$ cm$^{-3}$) and temperatures of $T\approx10^4$ K \citep[][]{Wiese1972} do not 
exhibit the blue-optical color temperature of a hot blackbody with $T\approx10^4$ K.  The necessary ingredient to build up optical depth
at this density and temperature, and hence produce 
a $T\approx10^4$ K \emph{blackbody-like} spectrum with a small Balmer
jump, is a large physical depth range ($\Delta z$) of emitting material along the line of sight.  
Thus, the color temperature in flare spectra does not directly relate to 
the atmospheric temperature so much as it is an indication of how the amount
of emitting material along the line of sight varies as a function of
wavelength (indeed, material must
be heated to $T\approx10^4$ K for hydrogen to be significantly
populated out of the ground state into $n=2$ and $n=3$ and generate optical depth).
In the CC of the F13 atmosphere, the physical depth range
over which continuum emission escapes is $\Delta z_{\rm{FWHM}} \approx 1-2$ km,
which is not present in laboratory experiments.  At the blue
continuum wavelengths with the lowest opacity (\emph{e.g.}, at
$\lambda=4300$\ \AA), the physical depth range over which continuum emission escapes is relatively large in the CC and extends
  to lower heights at $z\approx200$ km (50 km 
below the CC).  The amount of emitting material along a line of sight can also change due to variations in the 
density at the heights where continuum emission escapes; in the F13 flare atmosphere,
the density is smaller at $z\approx200$ km compared to the maximum density in the CC (Figure \ref{fig:dynamics}).  Thus, the integration of the \emph{height-dependent} properties (as for the contribution function in Figure \ref{fig:didz1}) is important for determining the emergent intensity
and flux spectra.  

Another important result is that by applying the
 occupational probability formulae \citep{HM88, Dappen1987, Tremblay2009} 
to NLTE b-f and b-b Balmer opacities, we can explain the pseudo-continuum properties in the spectral region redward
of the Balmer edge ($\lambda>3646$\ \AA) and between the Balmer lines
from $\lambda=3750-3900$\ \AA.  These formulae approximate how wavelengths redward of the Balmer jump
experience Balmer b-f opacity in addition to Paschen b-f opacity (and
f-f opacity); the
atomic processes that allow for this additional Balmer continuum opacity are 
dissolved upper levels of hydrogen from Stark perturbations by ambient
protons and Landau-Zener
transitions of electrons between dissolved levels of hydrogen.  The amount of Balmer continuum emission 
redward of the Balmer jump and the amount of high order b-b Balmer
emission vary inversely, and therefore the relative
amounts of dissolved and undissolved components in spectra provide a 
 new diagnostic of ambient charge density in the white-light emitting,
 partially ionized regions in the lower
flare atmosphere (Section \ref{sec:wn2}, Table \ref{table:wndec}).

The heating mechanism at high density in dMe flares is still unresolved since the CC required
 to produce the observed properties of the white-light emission results from a NT electron beam that would experience 
 a considerable loss of energy from the return-current electric field.
The main results of this work are therefore the new interpretation of the blue-optical color temperature and Balmer jump
ratio (as the variations in the physical depth range and optical depth as a function of wavelength) and of the apparent pseudo-continuum between the higher
order Balmer lines (as Balmer continuum emission due to Landau-Zener transitions of electrons between dissolved levels). 
 Combined with future modeling efforts using a more plausible heating mechanism (Section \ref{sec:work}), these insights will allow the spectral 
observables from individual flares to directly constrain the atmospheric properties in the dense, lower atmosphere.

\section{Future Modeling Improvements} \label{sec:work}

We plan to develop a grid of short flare bursts using a larger
low-energy cutoff of the NT electron spectrum, thereby reducing the
number flux of NT electrons (and thus return-current effects) while allowing a larger fraction of energy to
penetrate deeper in the atmosphere.  A possible alternative is to employ lower NT electron fluxes and explore possible re-acceleration mechanisms in the chromosphere in a similar way as done recently by \citep{Varady2014}.
Including additional energy deposition mechanisms, such as NT proton energy deposition \citep[\emph{e.g.}, a neutral beam simulation such as in][]{Karlicky2000} or a combination of 
energy deposition by other means \citep[\emph{e.g.},][]{Fletcher2008} may
also be important for producing a similar atmospheric response with a smaller flux of NT electrons. 
In addition, we plan to incorporate the
spatial development of a flare arcade of sequentially heated
footpoints (\emph{e.g.}, as in the observations of the Bastille Day solar flare from \cite{Qiu2010} or the cartoon two-ribbon model of the megaflare event in \cite{Kowalski2012})
in order to reproduce the observed timescales of dMe flares, which are longer than the 5~s burst modeled here.

\begin{figure}[H]
\begin{center}
\includegraphics[scale=0.6]{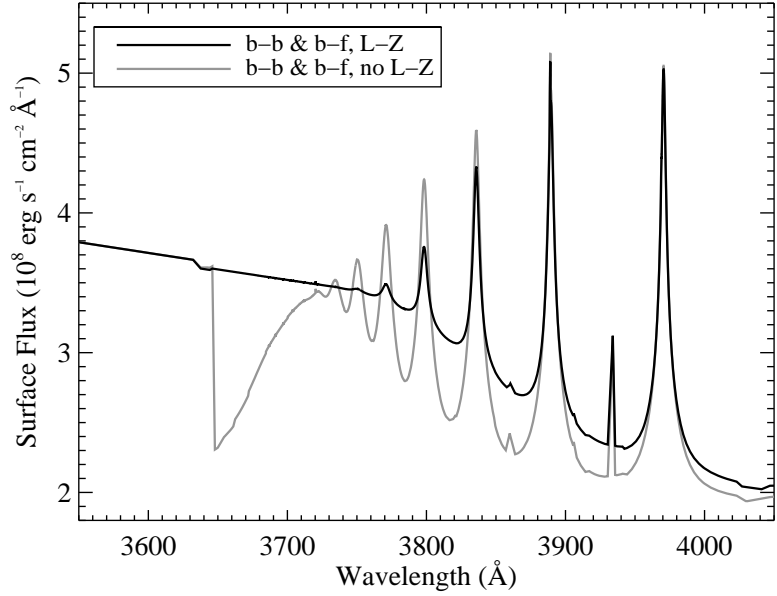}
\caption{ RH spectra of the Balmer jump region without instrumental
convolution applied and with a magnified flux scale.  \emph{Black line:} Electric field effects on b-b opacities and Landau-Zener
transition effects on b-b and b-f opacities from critical electric fields.  H11 is the bluest
 identifiable Balmer line.  \emph{Gray line:} 
Electric field effects on b-b opacities and Landau-Zener transitions not included. H13 is the bluest
identifiable Balmer line.
Including Landau-Zener transitions in the b-b and b-f opacities predicts more continuum emission between
3646\AA\ and 3720\ \AA, whereas not accounting for Landau-Zener transitions produces a ``blue continuum bump'' at $\lambda\approx3720$\ \AA\ from
the blending of Stark-broadened Balmer lines. Instrumental broadening is not shown; it broadens the
 Balmer lines but does not have a strong effect on the amount of 
 continuum between the lines.  }
\label{fig:pseudoC_S28_zoom}
\end{center}
\end{figure}

\begin{figure}[H]
\begin{center}
\includegraphics[scale=0.75]{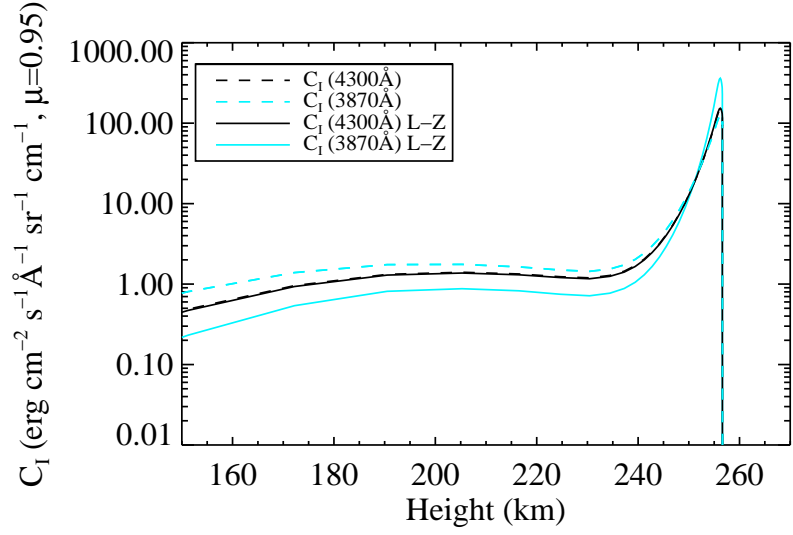}
\caption{ The contribution function at $\lambda=3870$\ \AA\ and 4300\ \AA\ accounting for opacity effects from Landau-Zener
transitions (solid) compared to the original calculation (dashed).  With L-Z transitions, the additional Balmer opacity at $\lambda=3870$\ \AA\ causes more emission to escape from the CC and less from the stationary flare layers than without the L-Z transitions.  The contribution function at $\lambda=4300$\ \AA\ is not affected.
}
\label{fig:contrib_cont_3870}
\end{center}
\end{figure}

\begin{figure}[H]
\begin{center}
\includegraphics[scale=0.6]{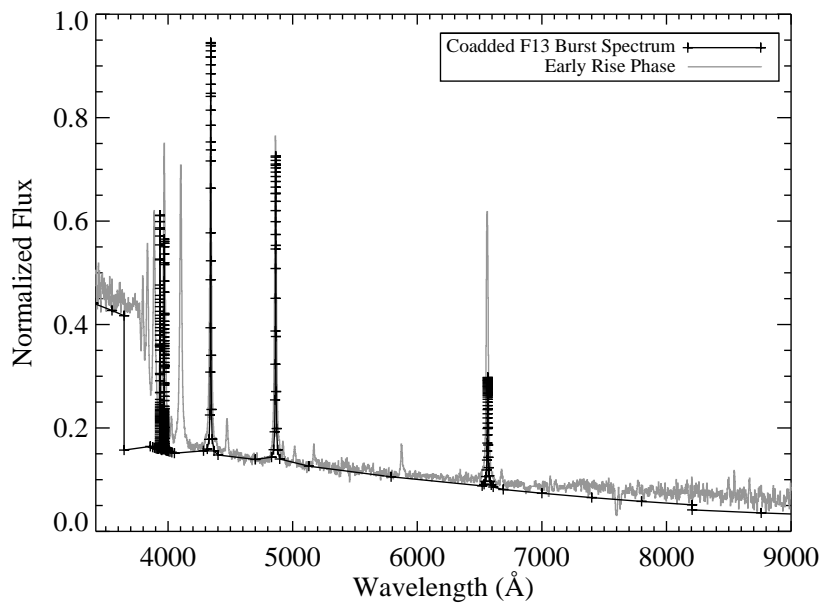}
\caption{The coadded (average) model F13 spectra (from RADYN; instrumental broadening is not applied) over 5~s compared to the early rise spectrum (S\#26)
during the IF3 flare from K13.  The coadded model spectrum has a larger 
Balmer jump and lower optical color temperature than the instantaneous model spectrum at $t=2.2$~s in Figure \ref{fig:pseudoC_S28}.
The flux at $\lambda < 3646$\ \AA\ increases towards bluer wavelengths in the model and observation.  Note that 
the opacity effects from Landau-Zener transitions are not included here because statistical equilibrium is a poor
approximation during the early time steps in the model (Section \ref{sec:wn}).
}
\label{fig:coadd}
\end{center}
\end{figure}

In future work with the RH code, we intend to improve the
implementation of occupational probability theory and Stark broadening.  
 \cite{Hubeny1994} fully extended the occupational probability formalism to NLTE, and
 we suggest that flare modeling codes developed in the future should similarly include occupational
probability formalism in the non-equilibrium rate equation \citep[following Equation 2.23
in][]{Hubeny1994} in addition to the NLTE b-b and b-f opacity and emissivity.   
Here, our goal was to demonstrate that large improvements in the NUV model spectrum result from implementing the
 simple modifications to the opacity and emissivity given in \cite{Dappen1987} and \cite{Tremblay2009}.  
We will also follow the work of \cite{Tremblay2009} and \cite{Seaton1990} and \emph{renormalize}
the Stark profiles of \cite{Vidal1973} using $\beta_{\mathrm{crit}}$
as an integration limit, which leads to significantly narrower 
higher order Balmer lines \citep[\emph{cf.} Figure 5 of ][]{Tremblay2009}
 The theory of \cite{Vidal1973} is only valid to first order when the 
line wings do not overlap, since the Landau-Zener continuum accounts for larger perturbations. Therefore, 
too much emission is probably produced between the Balmer lines between $\lambda=3800-4000$\ \AA\ in the model spectrum in
 Figure \ref{fig:pseudoC_S28_zoom}.  
Between $\lambda=3646-3800$\ \AA, the low values of $w_n$ (that multiply into the b-b opacity) lead to only 
a small amount of blended Balmer line wings compared 
 to the Landau-Zener continuum, so the error is not important.  
Finally, the linear Stark broadening theory of the Balmer lines using the analytic approximations
of \cite{Sutton1978} as Voigt damping parameters should be replaced by the full Stark profiles from \cite{Vidal1973} and \cite{Lemke1997} for the conditions 
present in flare atmospheres, in order to give accurate diagnostics on the electron density for unblended Balmer lines.  

In future work with the RADYN code (Allred \emph{et al.} 2015, in preparation), we intend to include the following
improvements to the modeled physics:  the cooling from Mg \textsc{ii} and Fe \textsc{ii} ions, 
an improved optically thin radiative loss function, non-thermal ionization of He \textsc{i} and He \textsc{ii}, return-current heating, 
magnetic mirroring of the NT electrons, and the fully relativistic Fokker-Planck solution
to energy deposition and pitch angle scattering of NT electrons.

\clearpage

\begin{table}
%\tabletypesize{\scriptsize}
%\rotate
%\tablewidth{0pt}
\caption{Observables calculated from the model flux spectra.  These values have been obtained without subtracting the preflare spectrum.}\label{table:observables}
\begin{tabular}{llll}
\hline
Model (time step; phase) &
$\chi_{\mathrm{flare}}$ &
$T_{\mathrm{BB}}$ (K) &
H$\gamma_{\rm{EW4170}}$ \\

\hline
F11 ($t=2.2$s; impulsive peak) & 9.2  & 5300 & 150 \\
F12 ($t=2.2$s; impulsive peak) & 7.4 & 5700 & 85 \\ % 88
F13 ($t=0.2$s; impulsive early, rise 1) & 6.0   & 6300 & 70  \\ % 68
F13 ($t=1.2$s; impulsive late, rise 2) & 3.3   & 7400 & 40  \\ % 38
F13 ($t=2.2$s; impulsive peak) & 2.0 & 9300 & 20 \\ % 19
F13 ($t=4$s; gradual decay) & 3.3 & 5400 & 60 \\
F13 (average burst) & 2.7 & 7500 & 30 \\

\end{tabular}
\end{table}

\begin{table}
\caption{The Landau-Zener Balmer decrement (H11/H$\gamma$) in the optically thick, isothermal, LTE limit.  Here, we give only the charged
component of $w_n$ (Equation (\ref{eq:wn}), averaged for the range of temperatures), 
ignoring the minor dependence on the destruction of levels from collisions with other neutrals.}\label{table:wndec}

\begin{tabular}{llllll}
\hline
$n_p$ (cm$^{-3}$) &
$w_{n=11}$&
$T = 10\,000$ K &
$T = 12\,500$ K &
$T = 15\,000$ K &
$T = 20\,000$ K \\
\hline
$10^{13}$        & 1.00 & 0.30 & 0.33 & 0.35 & 0.37 \\
$10^{14}$        & 0.98 & 0.30 & 0.32 & 0.34 & 0.37 \\
$5\times10^{14}$ & 0.85 & 0.26 & 0.28 & 0.30 & 0.32 \\ 
$10^{15}$        & 0.69 & 0.21 & 0.23 & 0.24 & 0.25 \\
$2.5\times10^{15}$ & 0.35 & 0.11 & 0.12 & 0.12 & 0.12 \\
$5\times10^{15}$ & 0.15 & 0.05 & 0.05 & 0.05 & 0.05 \\

\end{tabular}
\end{table}

%%%%%%%%%%%%%%%%%%%%%%%%%%%%%%%%%%%%%%%%%%%%%%%%%%%%%%%%%%%%%%%%%%%%%%%%%%%
%% Appendix
%
% \appendix   

%%%%%%%%%%%%%%%%%%%%%%%%%%%%%%%%%%%%%%%%%%%%%%%%%%%%%%%%%%%%%%%%%%%%%%%%%%%
%% Acknowledgements
%
 \begin{acks}
AFK thanks the Science Organizing Committee of the \emph{Solar and Stellar Flares} meeting in Prague, Czech Republic for the opportunity to present this work.  We thank
an anonymous referee for clarifications and comments that helped improve this work.
AFK thanks Dr. Petr Heinzel and Dr. Hans Ludwig for bringing hot star modeling papers of Stark broadening to his attention.  
We thank Drs. Adrian Daw, Eric Agol, Ellen Zweibel, and Jeremiah Murphy for helpful discussions and Dr. William Abbett for several
IDL routines used in the analysis.  AFK also acknowledges helpful discussions at the International Space Science Institute with 
Dr. Lyndsay Fletcher's Solar and Stellar Flares team and with 
Dr. Sven Wedemeyer-Bohm's Magnetic Activity of M-type Dwarf Stars and the Influence on Habitable Extra-solar Planets team.   
 This research was supported by an appointment to the
NASA Postdoctoral Program at the Goddard Space Flight Center, administered by Oak Ridge Associated Universities through
a contract with NASA, and by the University of Maryland Goddard Planetary Heliophysics Institute (GPHI) Task 132.
 \end{acks}

%%% %%%%%%%%%%%%%%%%%%%%%%%%%%%%%%%%%%%%%%%%%%%%%%%%%%%%%%%%%%%
%% Bibliography
%
% Using BibTeX
%

\clearpage
 \bibliographystyle{spr-mp-sola}
% %\bibliographystyle{spr-mp-sola-limited} %% Alternative style: no title
 \bibliography{kowalski_ms_v2}  
%
% Without BibTeX 
% \begin{thebibliography}{}
% \bibitem[\protect\citeauthoryear{Author}{Year}]{key}
%   <bibliographical entry>
%
% \bibitem[\protect\citeauthoryear{}{}]{}
%   
%  
% \end{thebibliography}

\end{article} 
\end{document}